\documentclass[
aps,
prx,
showpacs,
preprintnumbers,
twocolumn,
superscriptaddress,
]{revtex4-2}

\usepackage[T1]{fontenc}
\usepackage{newtxtext}
\usepackage{newtxmath}

\usepackage[normalem]{ulem}

\usepackage{enumitem}

\usepackage{makecell}

\usepackage{physics}
\usepackage{amsmath}
\usepackage{mathtools}
\usepackage{dsfont}

\usepackage{comment}

\usepackage{appendix}

\usepackage{placeins}

\usepackage{color}
\usepackage[dvipsnames]{xcolor}
\usepackage{colortbl}
\usepackage{hyperref}
\hypersetup{colorlinks,citecolor=blue,linkcolor=blue,urlcolor=blue}

\definecolor{darkgreen}{rgb}{0.55, 0.71, 0.00}
\definecolor{Gray}{gray}{0.9}

\usepackage{xpatch}

\makeatletter
\patchcmd{\@ssect@ltx}
{\addcontentsline{toc}{#1}{\protect\numberline{}#8}}
{}
{}
{}
\makeatother

\usepackage{xspace}

\usepackage{graphicx}
\usepackage[all]{hypcap}

\usepackage{tikz} % https://www.bu.edu/math/files/2013/08/tikzpgfmanual.pdf
\newcommand{\pluspole}{\raisebox{-3pt}{\tikz{\shade[ball color=red] (9,.5) circle [radius=0.15] node {{\color{white}$+$}};}}}
\newcommand{\minuspole}{\raisebox{-3.2pt}{\tikz{\shade[ball color=blue] (9,.5) circle [radius=0.15] node[below=-3.7pt]{{\color{white}$-$}};}}}

\graphicspath{{FiguresSI/}{FiguresMain/}}

\newcommand{\Hdd}{H_\mathrm{dd}}

\newcommand{\C}{$^{13}$C}

\newcommand{\app}{\approx}

\newcommand{\beq}{\begin{equation}}
\newcommand{\eeq}{\end{equation}}
                  
\newcommand{\benum}{\begin{enumerate}}
\newcommand{\eenum}{\end{enumerate}}
                    
\newcommand{\bit}{\begin{itemize}}
\newcommand{\eit}{\end{itemize}}
\newcommand{\xhat}{\hat{\T{x}}}

\newcommand{\bea}{\begin{eqnarray}}
\newcommand{\eea}{\end{eqnarray}}

\newcommand{\bev}{\begin{verbatim}}
\newcommand{\eev}{\end{verbatim}}

\newcommand{\qt}{\tau}

\newcommand{\lb}{\left(}
\newcommand{\rb}{\right)}

\newcommand{\T}[1]{\textbf{#1}}

\newcommand{\R}[1]{\textrm{#1}}
\newcommand{\ba}{\left\{ \begin{array}{lr}}
\newcommand{\ea}{\end{array}\right.}

\newcommand{\blist}[1]{
 \begin{list}{#1}%$\ast\circ\bullet\Right
 \begin{align}
	 arrow
 \end{align}
 $\checkmark\star
  { \setlength{\itemsep}{3pt}
     \setlength{\parsep}{2pt}
     \setlength{\topsep}{3pt}
     \setlength{\partopsep}{0pt}
     \setlength{\leftmargin}{1em}
     \setlength{\labelwidth}{1em}
     \setlength{\labelsep}{0.5em} } }
\newcommand{\elist}{
  \end{list}  }

\newcommand{\bef}
{
\begin{figure}[htbp]
\centering
}

\newcommand{\eef}{\end{figure}}

\newcommand{\beginmethods}{%
	\setcounter{table}{0}
	\renewcommand{\thetable}{M\arabic{table}}%
	\setcounter{figure}{0}
	\renewcommand{\thefigure}{M\arabic{figure}} %
	\renewcommand{\theHfigure}{M\arabic{figure}} %fixes linking to figures
	\renewcommand{\figurename}{Fig.} 
	\renewcommand{\thesection}{\:M\arabic{section}}
	\setcounter{section}{0}
	\setcounter{equation}{0}
	\renewcommand{\theequation}{M\,\arabic{equation}}
}
\newcommand{\beginsupplement}{%
	\setcounter{table}{0}
	\renewcommand{\tablename}{Supplementary Table}
	\renewcommand{\thetable}{\arabic{table}}%
	\setcounter{figure}{0}
	\renewcommand{\thefigure}{S\arabic{figure}} %
	\renewcommand{\theHfigure}{S\arabic{figure}} %fixes linking to figures
	\setcounter{page}{1}
	\renewcommand{\figurename}{Fig.} 
	\renewcommand{\thesection}{\:S\arabic{section}}
	\setcounter{section}{0}
	\setcounter{equation}{0}
	\renewcommand{\theequation}{S\,\arabic{equation}}
}

\newcommand{\affA}{Department of Chemistry, University of California, Berkeley, Berkeley, CA 94720, USA.}
\newcommand{\affB}{Max Planck Institute for the Physics of Complex Systems, N\"othnitzer Str.~38, 01187 Dresden, Germany.}
\newcommand{\affC}{Chemical Sciences Division,  Lawrence Berkeley National Laboratory,  Berkeley, CA 94720, USA.}

\newcommand{\affF}{CIFAR Azrieli Global Scholars Program, 661 University Ave, Toronto, ON M5G 1M1, Canada.}
\newcommand{\affG}{School of Physics, Peking University, 100871, Beijing  China}
\newcommand{\affH}{Department of Physics, Harvard University, Cambridge, MA 02138, USA}

\newcommand{\affK}{Technical University of Munich, TUM School of Natural Sciences, Physics Department, 85748 Garching, Germany}
\newcommand{\affL}{Munich Center for Quantum Science and Technology (MCQST), Schellingstr. 4, 80799 M{\"u}nchen, Germany}
\newcommand{\affM}{Blackett Laboratory, Imperial College London, London SW7 2AZ, United Kingdom}

\begin{document}
	
	%%%%%%%%%%%%%%%%%%%%%%%%%%%%%%%%%%%%%%%%%%%%%%%%%%%%%%%%%%%%%%%%%%%%%%%%%%%
	%                        Main Document
	%%%%%%%%%%%%%%%%%%%%%%%%%%%%%%%%%%%%%%%%%%%%%%%%%%%%%%%%%%%%%%%%%%%%%%%%%%%
	%%%%%%%%%%%%%%%%%%%%%%%%%%%%%%%%%%%%%%%%%%%%%%%%%%%%%%%%%%%%%%%%%%%%%%%%%%%

	\title{Experimental observation of a time rondeau crystal:\\ Temporal Disorder in Spatiotemporal Order}
	%\title{Experimental Observation of Temporal Disorder in Spatiotemporal Order} 
	
	\author{Leo~Joon~Il~Moon}
	\thanks{equal contribution}
	\affiliation{\affA}
	
	\author{Paul~M.~Schindler}
	\thanks{equal contribution}
	\affiliation{\affB}
	
	\author{Yizhe~Sun}
	\affiliation{\affB}
	\affiliation{\affH}
	
	\author{Emanuel~Druga}
	\affiliation{\affA}
	
	\author{Johannes~Knolle}
	\affiliation{\affK}
	\affiliation{\affL}
	\affiliation{\affM}
	
	\author{Roderich~Moessner}
	\affiliation{\affB}
	
	\author{Hongzheng~Zhao}
	\email{hzhao@pku.edu.cn}
	% \affiliation{\affB}
	\affiliation{\affG}
	
	\author{Marin~Bukov}
	\email{mgbukov@pks.mpg.de}
	\affiliation{\affB}
	
	\author{Ashok~Ajoy}
	\email{ashokaj@berkeley.edu}
	\affiliation{\affA}
	\affiliation{\affC}
	\affiliation{\affF}

	\begin{abstract}
		Our understanding of phases of matter relies on symmetry breaking, one example being water ice whose crystalline structure breaks the continuous translation symmetry of space. Recently, breaking of time translation symmetry was observed in systems not in thermal equilibrium. The associated notion of time crystallinity has led to a surge of interest, raising the question about the extent to which highly controllable quantum simulators can generate rich and tunable temporal orders, beyond the conventional classification of order in static systems.
		Here, we investigate different kinds of partial temporal orders, stabilized by non-periodic yet structured drives, which we call \textit{rondeau} order. 
		Using a \C-nuclear-spin diamond quantum simulator, we report the first experimental observation of a -- tunable degree of -- short-time disorder in a system exhibiting long-time stroboscopic order. 
		This is based on a novel spin control architecture that allows us to implement a family of drives ranging from structureless via structured random to quasiperiodic and periodic drives.
		Leveraging a high throughput read-out scheme, we continuously observe the spin polarization over $10^5$ pulses to probe rondeau order, with controllable lifetimes exceeding 4 seconds. 
		Using the freedom in the short-time temporal disorder of rondeau order, we show the capacity to encode information in the response of observables.
		Our work broadens the landscape of observed nonequilibrium temporal order, paving the way for new applications harnessing driven quantum matter.
	\end{abstract}
	
	\maketitle

	\begin{figure*}[t]
		\centering
		\includegraphics[width=\textwidth]{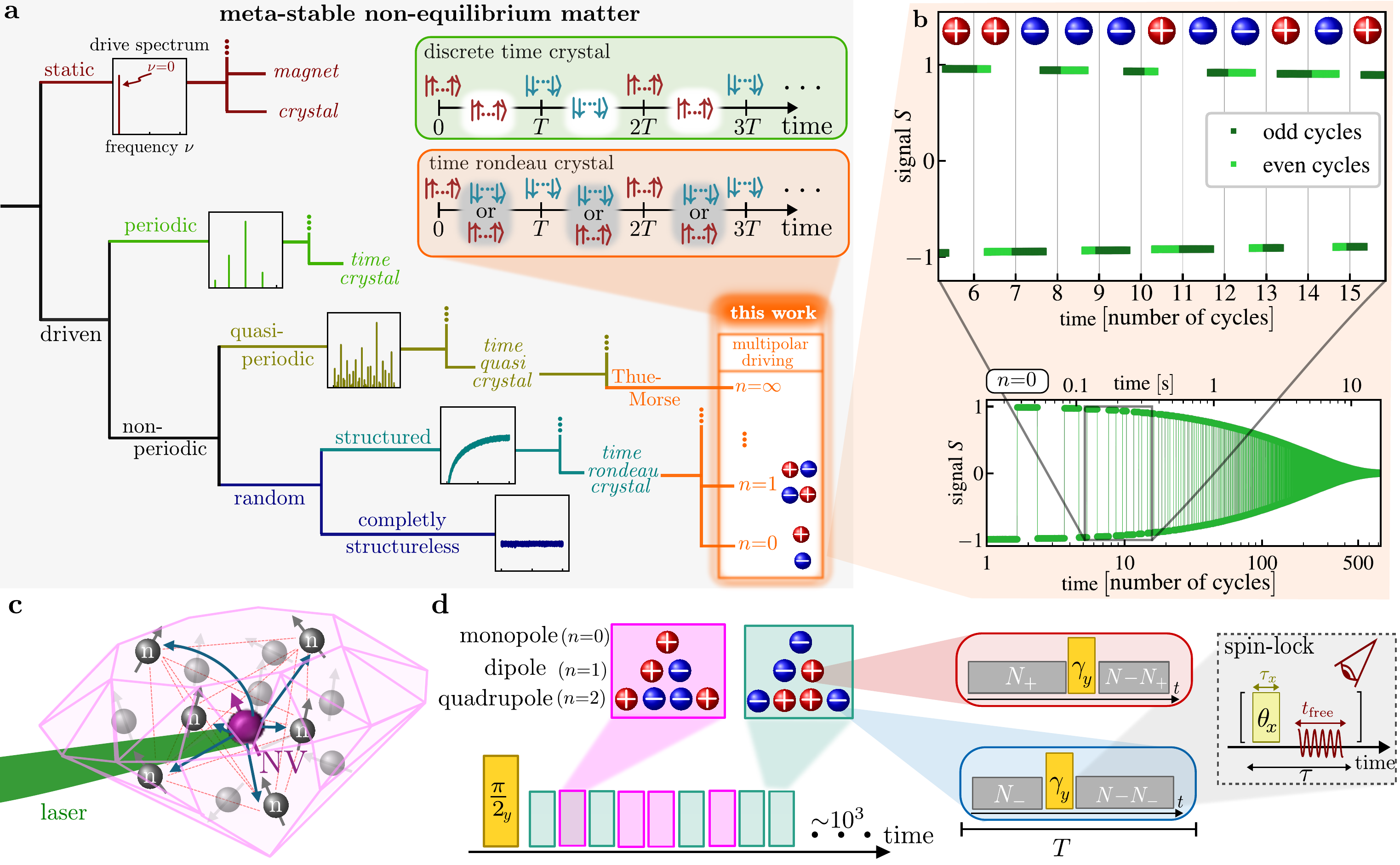}
		\caption{\textbf{Conceptualization, and experimental realization, of a time rondeau crystal:}
			\textbf{a}, Tree diagram giving an overview of parametrically (meta)stable equilibrium and nonequilibrium quantum matter (italics); tree branches show different types of drives and their spectral decomposition (square boxes). The family of random multipolar drives we implement interpolates between structured random and quasiperiodic drives (orange frame).
			Comparison of discrete time crystal~(green box) and time rondeau crystal~(orange box): while both temporal orders show period doubling dynamics at stroboscopic times, $t{=}MT$, the micromotion dynamics of the time rondeau crystal is (tunably) disordered.
			\textbf{b}, Exemplary data set of experimental observation of rondeau order. Data represents single shot measurements of the driven \C-nuclei with $\tau{=}74.4\,\mu\mathrm{s}$, $N_+{=}200$, $N_-{=}100$ at $\gamma_y{=}0.98\,\pi$ for a $0$-RMD sequence. \textit{Upper Panel:} Zoom into a window comprising $10$ full stroboscopic cycles. The signal flips sign after each full cycle; however, the point within one cycle where the signal flips is random, clearly indicating the coexistence of long-range temporal order and short-range temporal disorder. \textit{Lower Panel:} Full $16$s data set comprising $720$ pulses. The $1/e$ lifetime, $T_e$, exceeds $170$ periods, corresponding to $4\,\mathrm{s}$ or $JT_e {\approx} 2.6{\times} 10^3$.
			\textbf{c-d} Experimental implementation: 
			\textbf{c}, system comprising randomly placed dipolar interacting \C\ nuclear spins in diamond. Dashed lines indicate relevant dipole-dipole interactions. The nuclear spins are hyperpolarized by optically pumped NV centers.
			\textbf{d}, Experimental implementation of random multipolar driving (RMD) protocol: $n$-RMD sequence consists of randomly anti-aligning $n$-multipole sequences with equal probability (magenta/teal pulses). The $n$-multipole is constructed systematically from the two monopoles, \protect\pluspole~and \protect\minuspole (magenta/teal boxes), which are defined via a two-tone drive as follows (red/blue ovals). A spin-locking train, consisting of $N$ spin-lock pulses (grey box), is interrupted after $N_\pm$ pulses by a $y$-pulse of angle $\gamma_y$ (orange pulse).
		}
		%Comments:  
		\label{fig:intro}
	\end{figure*}

	\paragraph*{Introduction.}

	The quest to define notions of order and disorder as organising principles of the natural world is one of the oldest endeavours of science and philosophy. The existence of water in solid, liquid, and gaseous forms is a matter of everyday experience, but it also illustrates the complexity of such notions. While liquid and gas phases lose their sharp distinction at high pressure, the solid phase, ice, incorporates a high degree of disorder in the location of the protons bonding the oxygen ions: spatial order over long distances coexists with disorder at short lengthscales. 
	
	The modern theory of phases and phase transitions, associated with the names of Landau, Ginzburg, and Wilson, has the notion of symmetry and its breaking at its core \cite{chaikin_lubensky_book,legget2004nobel_lecture}; adding topological forms of order \cite{wen1990ground-state}, invisible to local symmetry, completes our current understanding.
	For instance, through its ordered crystalline lattice of oxygen ions, ice breaks the translational symmetry of space. Yet, breaking translational symmetry in time~\cite{wilczek2012quantum} is forbidden in thermal equilibrium~\cite{2013PhRvL.111g0402B,watanabe2015absence}.

	However, in settings where thermal equilibrium is not reached, a notion of `time crystallinity' can indeed be defined. 
	A prominent example is the discrete time crystal (DTC) in periodically driven (Floquet) systems~\cite{khemani2016phase,PhysRevLett.118.030401,sacha2017time,zhang2017observation,autti2018observation,pizzi2019period,khemani2019brief,Else2020,kyprianidis2021observation,frey2022realization,mi2022time,mcginley2022absolutely,xiang2024long,shinjo2024unveiling}: DTCs exhibit long-range order in both space and time. In arguably their most robust form, the many-body localised DTC \cite{khemani2016phase}, this spatiotemporal phenomenon is a manifestation of a new notion of order, known as eigenstate order \cite{huse2013localization,khemani2019brief}, which can appear in disordered and interacting quantum systems. 
	
	In this work, we extend the above two threads by devising new types of {\it partial} temporal orders, including one we christen \textit{rondeau} order: The rondeau is a pattern comprised of a repeating principal theme (corresponding here to stroboscopic long-time order) that alternates with one or more contrasting variation themes (short-time disorder)~\footnote{By its origin, the rondeau is a form of medieval and Renaissance French poetry, as well as the corresponding musical chanson form. Together with the ballade and the virelai it was considered one of three formes fixes, and one of the verse forms in France most commonly set to music between the late 13th and the 15th centuries. Perhaps one of the more famous examples of a rondeau in music is Mozart's \textit{Rondo alla Turca} (Turkish March) [from \href{https://en.wikipedia.org/wiki/Rondeau_(forme_fixe)}{Wikipedia}].}.
	Rondeau order combines temporal disorder on short timescales with temporal order on long ones. More precisely, stroboscopic order at special points of the drive cycle coexists with {-- a tunable degree of --} disorder at all other times. This is in stark contrast to the Floquet lore, where stroboscopic observations at arbitrary points of the drive cycle all yield essentially the same temporal order, and are distinguished only by the entirely regular so-called micromotion within a drive cycle. 
	
	Specifically, we provide a genealogy of {\it non-periodic}  but {\it structured} drives that form such new types of partial temporal order, shown in Fig.~\ref{fig:intro}a with their corresponding Fourier spectra. 
	They interpolate between structureless random drives, associated with full temporal disorder at the one end, and encompass deterministic quasiperiodic drives that can give rise to quasicrystalline temporal order~\cite{dumitrescu2018logarithmically,zhao2019floquet,else2020long,he2024experimental}, at the opposite extreme. 
	
	The key experimental aspect of this work is the explicit demonstration of the existence of such orders in a macroscopic system of interacting, hyperpolarized, \C-nuclear spins at room temperature [Fig.~\ref{fig:intro}b]. 
	Leveraging a new spin control architecture based on an arbitrary-waveform generator with extensive sequence memory (Methods), we are able to accurately implement stable long drive protocols that realize a wide spectrum of nonequilibrium time-dependent drives, including structureless and structured random, quasiperiodic, and periodic sequences. Such capability permits us to experimentally investigate random multipolar drives (RMDs)~\cite{RMDDTC_Zhao_etal2023} -- a family of structured random protocols with controllable heating channels, allowing for long spin polarization lifetimes. During the long-lived metastable (prethermal) regime, we continuously monitor the system exhibiting rigid stroboscopic oscillations of local observables (like Floquet DTCs), but with a temporally disordered micromotion [Fig.~\ref{fig:intro}d].
	
	For a family of structured RMDs, including the quasiperiodic Thue-Morse sequence, we observe robust stroboscopic DTC order across a large parameter range, with long prethermal lifetimes comprising well over one hundred~(${\sim}170$) cycles, corresponding to $>4\,\mathrm{s}$.
	
	The Fourier spectrum of the micromotion reveals distinctive dynamical features of the time rondeau crystal compared to ordinary DTCs: randomness in time traces of observables gives rise to a smooth Fourier spectrum in stark contrast to isolated peaks observed for periodic or quasiperiodic drives [Fig.~\ref{fig:intro}a, insets].
	We demonstrate a parametrically controlled lifetime of the prethermal temporal order, by changing the drive period and analyzing imperfections of the applied pulses. We find little to no dependence of the lifetime on the details of the drive, in agreement with numerical simulations and analytical predictions.
	This enables us to freely engineer the form of the nondeterministic micromotion dynamics, without compromising the stability of the coexisting long-range temporal DTC order.

	\paragraph*{System.}
	
	Our experiment is performed at room temperature on a single-crystal diamond doped with ${\sim} 1 \,\mathrm{ppm}$ NV centres, and hosting a natural abundance ($1.1\%$) of \C\ nuclei. We utilize the randomly positioned network of \C\ nuclear spins, described by the Hamiltonian
	\begin{equation}\label{eq:dip}
		\Hdd =  \sum_{k<l} B_{kl} \left[ 3 I_k^z I_l^z - \boldsymbol{I}_k \cdot \boldsymbol{I}_l \right],
	\end{equation}
	where $\boldsymbol{I}_{k}$ is the vector of spin-1/2 operators for nuclear spin $k$, and the long-range dipole-dipole interaction decays as $B_{kl}{\propto} 1/r_{kl}^3$, with $r_{kl}$ being the distance between two spins. The interaction strength between \C\ nuclear spins can be characterized by their median coupling $J{=}\langle B_{kl} \rangle{=}0.66\,\mathrm{kHz}$~\cite{Beatrez21_90s}. By applying a chirped microwave drive, we transfer polarization from optically-pumped NV centers to the \C\ nuclear spins, hyperpolarizing them. The density matrix of the initial state is then $\rho_0 {\sim} \mu I^z$ ($I^{z}{=}\sum_{k}I_{k}^{z}$) with $z$-polarization fraction $\mu{\approx} 1\,\%$, enhanced ${\sim}998$-fold over its thermal equilibrium value~\cite{ajoy2020dynamical}.
	
	The experimental sequence is detailed in Fig.~\ref{fig:intro}c: after $60$~s of hyperpolarization~(Methods) the spins are tipped onto the $x$-$y$ plane by applying a $(\frac{\pi}{2})_{y}$ pulse. We then deploy a two-tone driving scheme consisting of a `fast' and a `slow' drive. 
	The `fast' drive comprises short spin-lock $(\frac{\pi}{2})_{x}$ pulses (length $\tau_x$), interleaved with free evolution governed by $H_\text{dd}$ for a time $t_\text{free}$; the duration of one spin-lock cycle is $\tau {=} \tau_{x} {+} t_\text{free}$ [Fig.~\ref{fig:intro}c, grey spin-lock box]. We measure the polarization of the spins in the $x$-$y$ plane inductively through an RF coil during the free evolution after each $x$-pulse~(Methods); such readout scheme allows us to track the polarization of the spins non-destructively through repeated weak measurements~\cite{DTC_Beatrez2022}. As a result, spin evolution can be traced over long times and hundreds of thousands of pulses quasi-continuously and in a single shot, a unique feature of our experiments compared to other quantum simulation platforms~\cite{rovny2018observation,choi2017observation,randall2021many}. Applying the `fast' drive imprints an emergent $U(1)$-symmetry associated with the conservation of $I^x$-polarization; this quasi-conservation law enhances the lifetime of the $x$-polarized initial state by over four orders of magnitude (compared to the bare nuclear $T_2^{\ast}{\app}1/J{=}1.5\,$ms)~\cite{Beatrez21_90s,DTC_Beatrez2022} (Methods).

	Temporal DTC order is realized using the `slow' drive, which consists of $y$-pulses of angle $\gamma_y{=}\pi{+}\varepsilon$~(length $\tau_y$, orange blocks in inset of Fig.~\ref{fig:intro}c) with free evolution time $t_{\text{free}}$ after every $\gamma_{y}$ pulse, interspersed between trains of `fast' spin-lock cycles (grey blocks) at multiples of $\tau$.
	To implement a structured random drive, we use RMD following the proposal of Ref.~\cite{RMDDTC_Zhao_etal2023}. We define two RMD \textit{monopoles}~($n{=}0$) as follows: the \pluspole/\minuspole sequence consists of two spin-lock trains of $N_{+/-}$ and $N{-}N_{+/-}$ spin-lock cycles respectively, with a single $y$-pulse in between [red/blue ovals, Fig.~\ref{fig:intro}d]; both sequences have the same total duration of $T$.
	Higher-order $n$-multipole pairs, e.g., dipoles ($n{=}1$: \pluspole\minuspole, \minuspole\pluspole), quadrupoles ($n{=}2$: \pluspole\minuspole\minuspole\pluspole, \minuspole\pluspole\pluspole\minuspole), etc., can be recursively constructed by anti-aligning $n{-}1$-multipole pairs together [magenta/teal boxes in Fig.~\ref{fig:intro}c]; the $n{\to}\infty$ limit corresponds to a deterministic quasiperiodic Thue-Morse drive \cite{allouche1999ubiquitous,nandy2017aperiodically,RMD_Rigorous_Mori_etal2021}.
	Finally, for a fixed $n$, the complete structured RMD is built out of placing the two sequences of a multipole pair randomly in time [Fig.~\ref{fig:intro}c, magenta/teal pulses in main protocol] (Methods). Although it is disordered as we illustrate below, the characteristic timescale $T$ remains fixed and hence defines a "period" for the drive; we refer to times integer multiple of $T$ as stroboscopic, $t{=}MT$, with $M{\in}\mathbb{N}$ the cycle number, and to all other times -- as micromotion.

	\begin{figure}[t]                  
		\centering
		\includegraphics[width=0.5\textwidth]{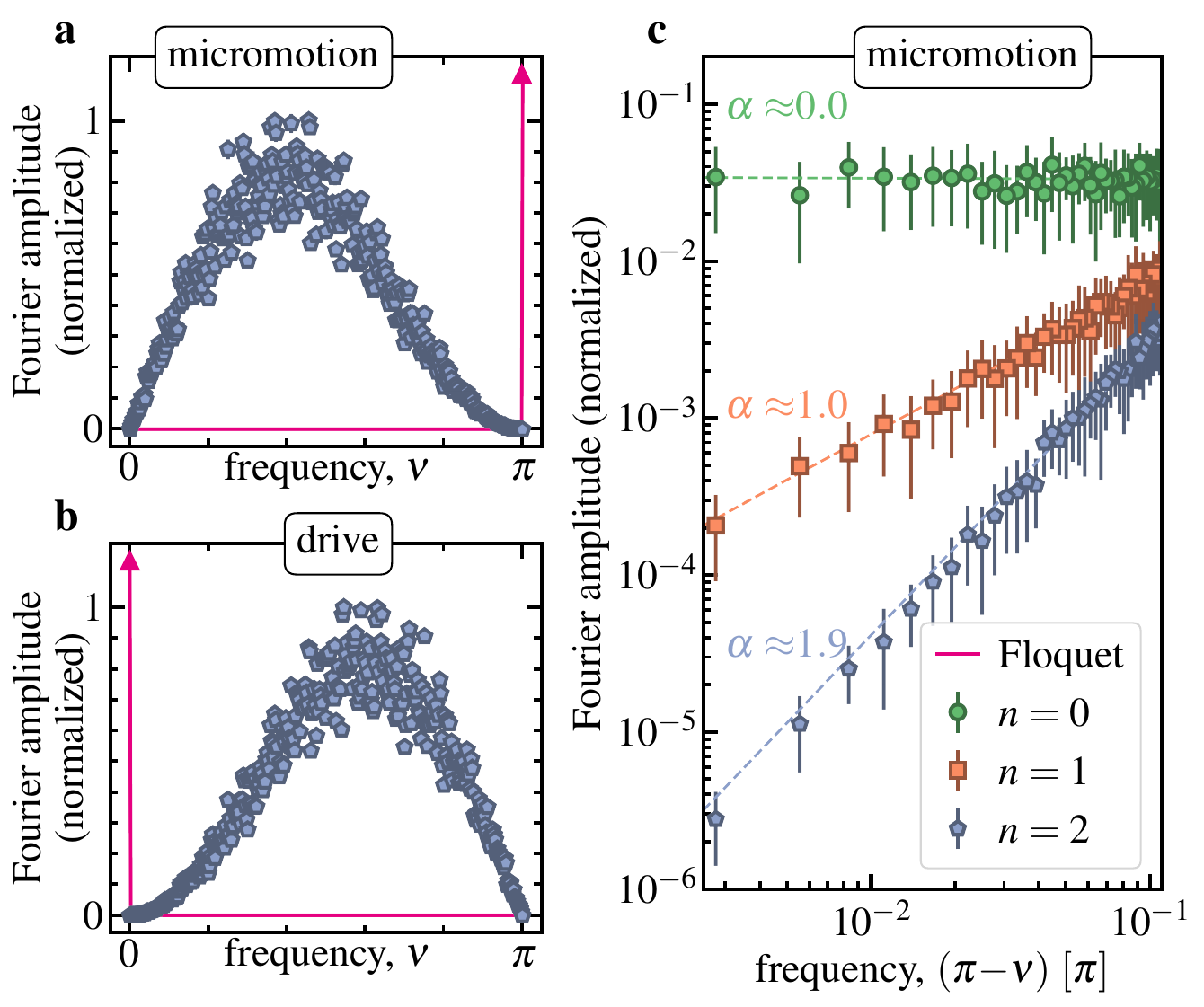}
		\caption{
			\textbf{Characteristics of rondeau order:}
			\textbf{a}, Amplitude of discrete Fourier transform~(DFT) of the polarization micromotion dynamics for Floquet DTC~(pink) and $2$-RMD~(blue). In contrast to Floquet DTCs, which feature only a single delta peak, the micromotion of the time rondeau crystal has finite support on the entire frequency spectrum.
			\textbf{b}, Amplitude of DFT of the $2$-RMD sequence that generated the data in a. The Fourier amplitudes of the drive and micromotion are mirror images of one another w.r.t.~the $\nu{=}\pi/2$ axis (referred to as $\pi$-shifted).
			\textbf{c}, Fourier amplitudes for RMD sequences $n{=}0,1,2$ on a log-log scale. The dashed lines are linear fits to the data with slopes $\alpha_0{=}(0.0{\pm} 0.1)$, $\alpha_1{=}(1.0{\pm}0.1)$, and $\alpha_2{=}(1.9{\pm}0.1)$ for $n{=}0,1,2$, respectively. 
			The different multipole orders show distinctive high-frequency suppression in good agreement with the theoretical predictions $\alpha_n{=}n$.
			Each RMD data point is averaged over $20$ realizations of the random drive; error bars indicate standard deviation; we set $\varepsilon{=}0.03\pi$; other parameters are as in Fig.~\ref{fig:intro}.
		}
		%Comments:
		\label{fig:Floquet_vs_RMD}
	\end{figure}

	\paragraph*{Characteristics of rondeau order.}
	
	We begin by analyzing the two-tone drive in the fine-tuned case $\gamma_y{=}\pi$, where each $y$-pulse fully inverts the polarization $I^x {\to} - I^x$. In between two $y$-pulses, the $x$-polarization is protected due to the emergent $U(1)$-symmetry irrespective of its sign. Since both monopoles include exactly one $y$-pulse, the system flips its polarization deterministically with period $2T$, irrespective of the specific choice of monopole pairs. Hence, it establishes long-range temporal order like conventional DTCs. By contrast, micromotion dynamics at non-stroboscopic times ($N_+\tau {<} (t\mod T) {<} N_{-}\tau$) inherits randomness from the spin-flip operation: the polarization, which either flips sign or remains unchanged, depends on the monopole that has been applied. 
	
	In our experiment, we first confirm the coexistence of temporal order and tunable micromotion disorder for $n{=}0$, and show that they persist away from the fine-tuned limit. We introduce imperfect polarization inversions by moving away from the fine-tuned point, $\gamma_y{=}\pi{+}\varepsilon$, by a small but finite deviation~($\varepsilon{\neq} 0$). As shown in Fig.~\ref{fig:intro}d, the system exhibits stroboscopic period-doubling behavior together with a disordered micromotion for exceptionally long times: the signal persists even after $M{>}500$ monopole sequences ($Jt{\approx} 10^4$), corresponding to a physical lifetime of more than $10\, \mathrm{s}$. We also observe a similar dynamical behavior for the entire family of $n$-RMD protocols~[SI].
	
	The discrete Fourier transform (DFT) of the polarization micromotion (Fourier frequency $\nu$) allows us to experimentally verify the characteristic features of RMDs~\footnote{To remove the influence of prethermal decay on the Fourier transform, we consider the digitized signal of the micromotion: positive~(negative) values of the signal $S>0$~($S<0$) are mapped to $+1$($-1$).}. While for $n{=}0$, the DFT spectrum is flat since the drive and micromotion are both temporally disordered, for a $2$-RMD the DFT spectrum is structured and smoothly distributed over all frequencies, cf.~Fig.~\ref{fig:Floquet_vs_RMD}a. The spectrum of the micromotion signal reflects closely that of the $2$-RMD used to generate it. The two spectra are shifted with respect to each other [Fig.~\ref{fig:Floquet_vs_RMD}a,b], since the polarization inversion introduces a $(-1)^M$ phase between the signal and the drive, resulting in a $\pi$-shifted DFT~\cite{RMDDTC_Zhao_etal2023}.  
	This feature is difficult to observe for $n{=}0$ since both spectra are trivially flat; however, for more structured $n$-RMDs with $n{\geq} 1$, the multipolar correlation imprints in their characteristic spectrum an algebraic suppression $\nu^n$ for $\nu{\to} 0$~(Methods). This suppression shifts to $\nu{\to}\pi$ in the DFT of the micromotion, making the distinctive $\pi$-shift feature experimentally measurable.
	
	To test the predicted frequency law, we expose the system to an $n$-RMD with $n{=}0,1,2$, and plot the DFT spectrum of the micromotion signal on a log-log plot against a $\pi$-shifted $\nu$-axis, cf.~Fig.~\ref{fig:Floquet_vs_RMD}c. The three data sets can be fitted to a good agreement by straight lines, and confirm the anticipated $(\pi{-}\nu)^n$ scaling behavior.
	This behavior of the micromotion DFT in the time rondeau crystal comes in stark contrast to Floquet DTCs, where the micromotion is trivially period-doubled as is the stroboscopic dynamics, leading to a single delta peak in the corresponding DFT spectrum (pink arrow, Fig.~\ref{fig:Floquet_vs_RMD} a,b). Such a comparison thus serves as a smoking gun for observing novel types of temporal order beyond the conventional Floquet DTC paradigm~\cite{RMDDTC_Zhao_etal2023}. Conceptually, the above analysis shows that the generalization of temporal order can be conveniently understood in Fourier space.
	
	\begin{figure}[t]
		\centering
		\includegraphics[width=0.5\textwidth]{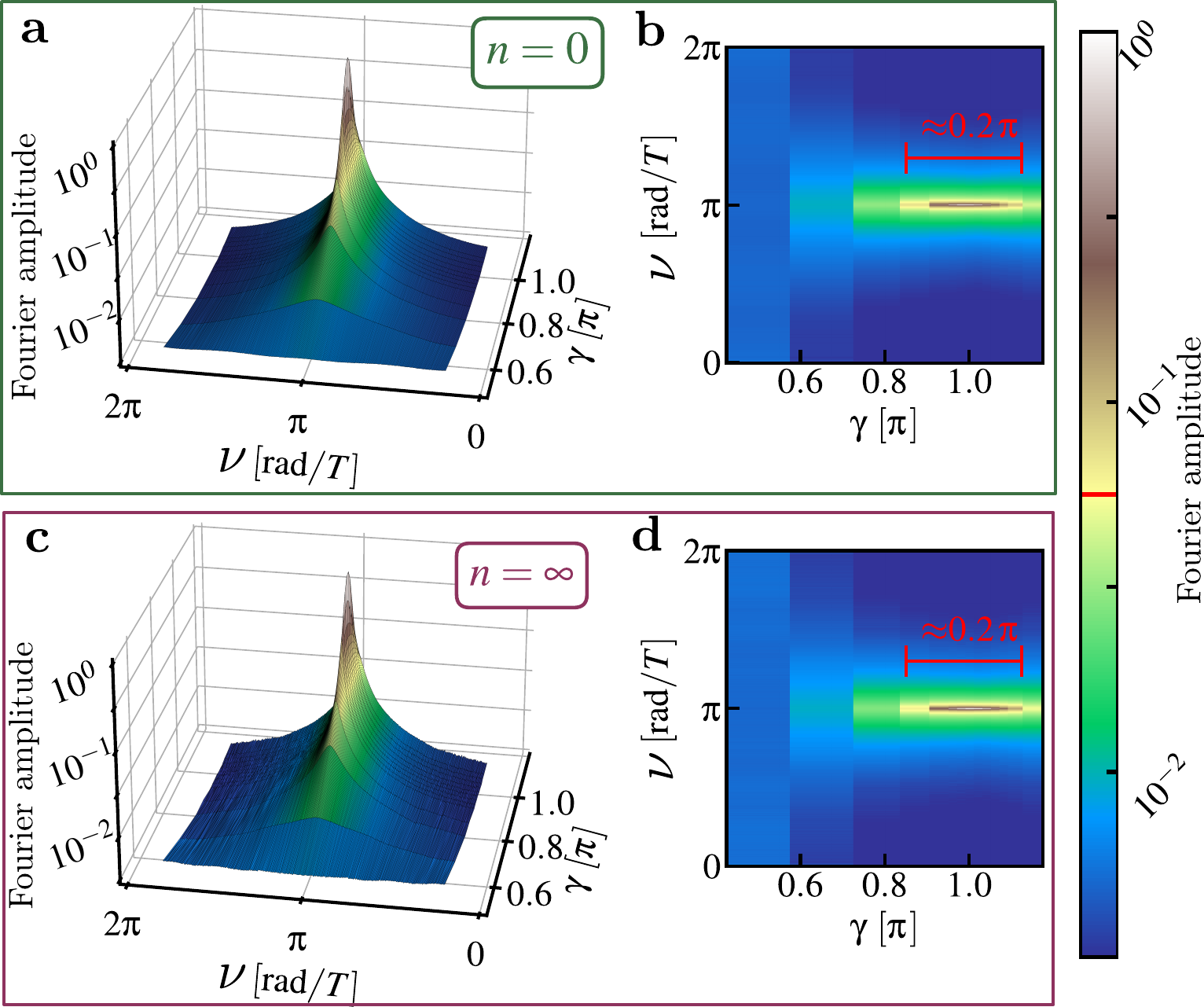}
		\caption{
			\textbf{Occurrence of prethermal rondeau order:}
			Normalized Fourier amplitudes of stroboscopic dynamics as a function of flip angle $\gamma_y$ around $\gamma_y{=}\pi$ and frequency $\nu$.
			\textbf{a}, Fourier intensities for monopole~($n{=}0$) sequence, and \textbf{b}, 2D projection.
			\textbf{c} and \textbf{d}, same as a and b but for the deterministic Thue-Morse~($n{=}\infty$) sequence.
			The prethermal temporal order shows a stable stroboscopic period doubling response over a large parameter regime, as indicated by the strong response of the Fourier intensity at half the drive frequency. Remarkably, the stability of the stroboscopic temporal order is independent of the RMD order $n$~(see SI for $n{=}1,2$).
			% Parameters
			Experimental $0$-RMD data is averaged over $10$ drive realizations. Other parameters are as in Fig.~\ref{fig:Floquet_vs_RMD}.
		}
		%Comments:
		\label{fig:phase_diagram}
	\end{figure}
	
	\begin{figure}[t]
		\centering
		\includegraphics[width=0.5\textwidth]{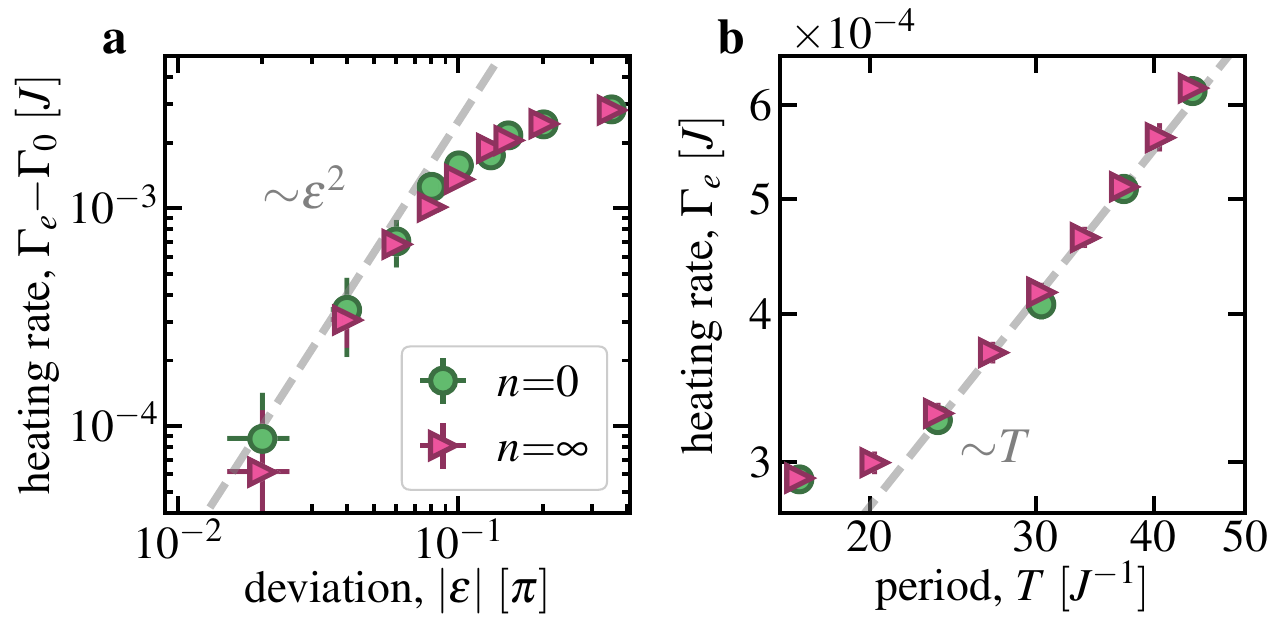}
		\caption{
			\textbf{Controllable prethermal heating rate.}
			\textbf{a}, dependence of heating rate $\Gamma_e {-} \Gamma_0$~($\Gamma_0{=}\Gamma_e{\mid}_{\varepsilon{=}0}$), on deviation $\varepsilon {=} \gamma {-} \pi$ for monopole~($n {=} 0$, green circles) and Thue-Morse~($n{=}\infty$, pink triangles) sequence; dashed line indicates power law ${\propto} \varepsilon^2$ predicted by numerical simulations [SI].
			\textbf{b}, heating rate $\Gamma_e$ against period $T$, while simultaneously changing the deviation $\varepsilon$ linearly in the period $\varepsilon {=} BT$ to keep the ratio $\varepsilon/JT$ fixed; we choose $B{/}(J\pi){\approx}5.9\times 10^{-4}$, other parameters are as in a; dashed line indicates power law ${\propto} T^1$ predicted by simulations.
			The nonequilibrium heating processes are systematically suppressed with decreasing deviation $\varepsilon$ and period $N\tau$, across the entire family of multipole orders~(see SI for $n{=}1,2$). Experimental $0$-RMD data are averaged over 20 drive realizations. The experimental data are consistent with power-law suppression of heating as ${\propto} \varepsilon^2$ and ${\propto} T$, respectively, as predicted in the dephasing limit~[SI]. 
		}
		%Comments:
		\label{fig:heating_rate}
	\end{figure}
	
	\paragraph*{Stability and Robustness.} 
	
	Since the time rondeau crystal is metastable and eventually melts, it is essential to analyze its lifetime and stability. The versatility of the driving protocols we implement allows us to efficiently scan over large parameter regimes and test the rigidity of the stabilized order against perturbations; we also have the ability to tune its lifetime parametrically over a large time window, as we now demonstrate.
	
	Notice first that the long-range temporal order and the short-range disorder actually share the same prethermal timescale, as can be seen from the long-time behavior of the polarization dynamics in Fig.~\ref{fig:intro}d; let us, therefore, focus on the stroboscopic dynamics. 
	To map out the phase diagram of the time rondeau crystal, we repeat the experiment for different values of the deviation parameter $\varepsilon$ from the perfect kick angle $\pi$, $\gamma_y{=}\pi{+}\varepsilon$, keeping the period $T$ fixed. We then calculate the DFT spectrum of the stroboscopic dynamics obtained from a fixed but long time window that comprises 720 pulses. 
	Fig.~\ref{fig:phase_diagram}a,b show the stroboscopic Fourier spectrum for a wide range of kick angles $0.5\pi \lessapprox \gamma_y\lessapprox 1.15\pi$, for a $0-$RMD drive. We find a dominant narrow peak centered around half the frequency~($\nu{=}\pi/T$) which spans over a finite range of kick angles ($\abs{\varepsilon}{\lessapprox} 0.1\,\pi$). This confirms the rigidity of rondeau order against small perturbations within this sufficiently long time window. For larger perturbations, the peak gradually fades away, suggesting a cross-over from rondeau to trivial order. 
	These experimental results are in excellent quantitative agreement with numerical simulations [SI].
	Remarkably, we observe similar behavior for other multipolar orders: $n{=}0,1,2,\infty$, see SI; hence, we find that rondeau order is robust across the entire family of RMD protocols. 
	In particular, the Thue-Morse sequence ($n{\to}\infty$) allows us to experimentally observe a robust prethermal time quasicrystal [Fig.~\ref{fig:phase_diagram}c,d], which shows the suitability of our experimental platform to explore a wide range of temporal orders across nonequilibrium matter [Fig.~\ref{fig:intro}a].

	Next, we quantify the parametric dependence of the decay rate $\Gamma_e$ of the spin polarization against changes in the period $T$ and the deviation $\varepsilon$ in the kick angle. We define $\Gamma_e$ as the inverse $1/e$ lifetime of the polarization, i.e., the time when the absolute value of the signal first drops below $1/e$ of its initial value. Note that, even for perfect kicks ($\varepsilon{=}0$), the polarization can still decay at a rate $\Gamma_0$ due to the approximate character of the emergent U(1) quasi-conservation law; $\Gamma_0$ can be systematically suppressed by, for instance, increasing the number of spin-lock trains $N$ per monopole~\cite{Beatrez21_90s}. Our data shows that deviations from this limiting case enhance the decay quadratically, $\Gamma_e{-}\Gamma_0{\propto} \varepsilon^2$, see Fig.~\ref{fig:heating_rate}a. 
	At the same time, we observe a linear suppression of $\Gamma_e$ for small $JT$ shown in Fig.~\ref{fig:heating_rate}b, regardless of the multipolar order $n$ used in the drive (a deviation for $JT{\ll}1$ is observed due to uncertainty in the calibration of the $\gamma_y$ pulse, see SI). 
	Both decay laws match our numerical simulations (grey dashed) with good accuracy, which can also be analytically justified by modeling the dynamics in a dephasing limit, see SI. 
	
	\paragraph*{Micromotion Engineering.}
	
\begin{figure}[t]
		\centering
		\includegraphics[width=0.5\textwidth]{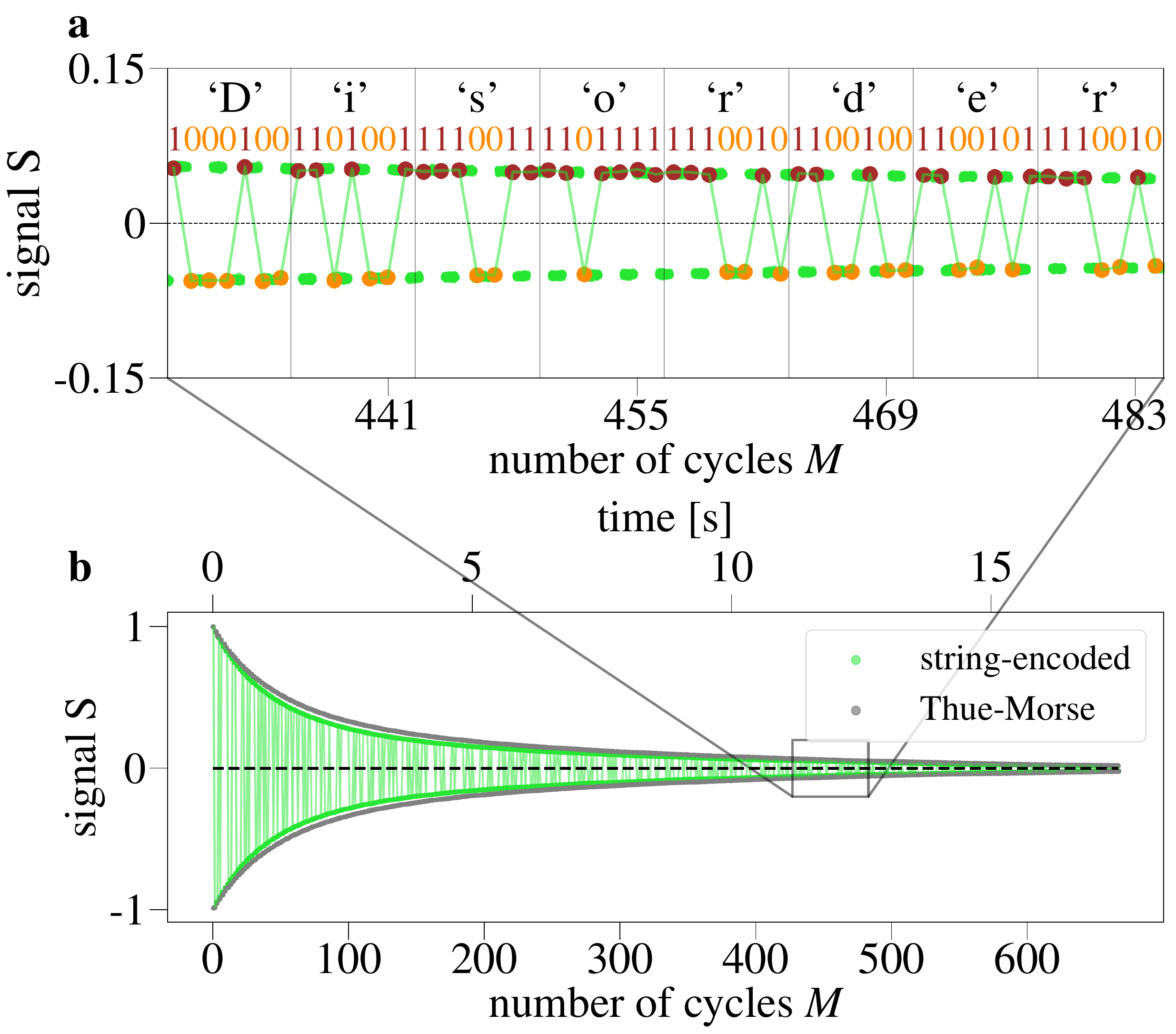}
		\caption{\textbf{\textit{`Experimental observation of a time rondeau crystal:\\ Temporal Disorder in Spatiotemporal Order'} encoded inside the micromotion of a time rondeau crystal} using ASCII encoding scheme. \textbf{a,} micromotion that encodes the word "Disorder" in ASCII encoding scheme is shown in full detail: `D' corresponds to $1000100$, `i' corresponds to $1101001$, etc. Light-green lines are $x$-projection signals of \C\ nuclear spin polarization measured at \textit{half-integer} periods [see text]. Brown (orange) dots indicate that the micromotion is positive (negative) and that the bit encoded is 1 (0). We can artificially engineer the order of \protect\pluspole/\protect\minuspole pulses to encode arbitrary information into the micromotion, as the heating is well suppressed even for a structureless random drive.
			\textbf{b,}
			Full dataset of $x$-projection signal of the string-encoding sequence~(green). The lifetime of the engineered micromotion is comparable to the lifetime of the quasiperiodic Thue-Morse sequence~(grey), demonstrating full controllability of micromotion without reducing the lifetime of the signal. Here $\gamma_{y}{=}\pi$ and $\theta_{x}{=}\pi/2$. 
			\label{fig:info_enc}
		}
	\end{figure}

	The controllably long lifetimes of rondeau order suggest potential for diverse applications. Specifically, in Fig.~\ref{fig:info_enc} we demonstrate versatile micromotion engineering for data encoding, using the sign of the rigid $x$-polarization values $+$ and $-$ as classical states to represent bits 0 and 1. We encode binary information in the sequential arrangement of the two monopole drives \pluspole and \minuspole. Since $N_+{>}N_{-}$, measuring the micromotion of the $x$-polarization at half-integer periods $(2\ell+1) T/2$ ($\ell{\in}\mathbb{N}$) is sensitive to polarization values before/after the $\gamma_y$-kick for the \pluspole/\minuspole unitary; this allows us to read-off the sign of the corresponding monopole. This paper's title can then be encoded within the micromotion dynamics of a time rondeau crystal in a 7-bit encoding system [Fig.~\ref{fig:info_enc}a]. In the experiment, which operates in the dephasing limit [SI], we find that the heating rate of the string-encoded drive aligns closely with the Thue-Morse sequence [Fig.~\ref{fig:info_enc}b]. Hence, we can fully manipulate the micromotion without compromising the lifetime of the underlying spatiotemporal order~\footnote{ 
		Compared to the $0$-RMD, encoding specific information necessitates a deterministic drive. The corresponding prethermal lifetime of the polarization may depend on the amount of structure (i.e., information) in the micromotion. We leave this open question for future exploration. 
	}.
	
	While Fig.~\ref{fig:info_enc} represents a proof-of-concept, the message size encoded can be significantly longer, continuing as long as the signal remains above the noise floor, here for $t{\app}36.2$s. Moreover, the continuous measurements can facilitate encoding akin to pulse code modulation~\cite{lathi2010modern}, enhancing capacity. Assuming a clock cycle of $\qt{=}86.8\mu$s, we estimate the ability to encode ${>}1.9{\times} 10^2$ characters (see SI), offering interesting new avenues for spatiotemporal memories~\cite{o2022random,guo2021efficient}.
	
	\paragraph*{Conclusion and outlook.}
	
	Our experiments open a promising new avenue to investigate temporal order, demonstrating, for the first time, long-lived stable coexistence of long-range temporal order and micromotion disorder at short timescales. Going beyond state-of-the-art techniques for controlling and probing nonequilibrium quantum matter, we are able to identify and implement random structureless and structured, quasiperiodic, and periodic drives that give rise to a wide range of temporal orders, including time-crystalline, time quasi-crystalline, and rondeau order -- all in a single quantum simulation platform. The versatile structure of our drive protocol allows us to map out the stability diagram and explore the robustness against external perturbations.

	Unlike ordinary DTC order, rondeau order allows for great tunability of the temporal spectral micromotion response, at a moderate cost on the lifetime of the temporal order. In fact, in our experiments, we observe no dependence of the lifetime on the details of the driving sequence. 
	Therefore, we can engineer arbitrary micromotion dynamics, beyond the RMD sequences considered before, without loss in signal quality, see e.g., Fig~\ref{fig:info_enc}.
	This enhanced tunability can boost potential applications of temporal order, like quantum sensing~\cite{iemini2023floquet, sahin2022high}, cat state preparation~\cite{bao2024schr}, or topological transport~\cite{simula2024topological}. 
	Specifically, the tunability of the power spectrum in our experiment may facilitate the creation of frequency-selective, DTC-based quantum sensors. Moreover, spin-lock lifetimes here are influenced by relaxation from NV electrons~\cite{Beatrez2023electron}; we instead anticipate significantly longer rondeau lifetimes in alternate systems based on photoexcited triplet electrons~\cite{eichhorn2014proton,singh2024room}. 
	Finally, although our experiment focused on nuclear spins in diamond, the underlying concept is immediately applicable to a wide swathe of quantum simulator platforms.

	\paragraph*{Acknowledgements.}
	
	We acknowledge technical contributions from J.~Ball, J.~Mercede (Tabor Electronics), and W.~Beatrez.
	We thank C.~Fleckenstein for many useful discussions in related previous collaborations in the beginning of this project. We also thank B.~Dou\c cot for pointing out the \textit{rondeau} analogy to us. 
	P.~M.~S.~thanks the Ajoy lab for their generous hospitality. 
	This work was funded by ONR (N00014-20-1-2806), AFOSR YIP (FA9550-23-1-0106), AFOSR DURIP (FA9550-22-1-0156), DNN NNSA (FY24-LB-PD3Ta-P38), the CIFAR Azrieli Foundation (GS23-013),  by
	the Deutsche Forschungsgemeinschaft  under cluster of excellence
	ct.qmat (EXC 2147, Project-ID No. 390858490),
	and by the European Union (ERC, QuSimCtrl, 101113633). Views and opinions expressed are however those of the authors only and do not necessarily reflect those of the European Union or the European Research Council Executive Agency. Neither the European Union nor the granting authority can be held responsible for them.
	Numerical simulations were performed on the MPIPKS HPC cluster.

	\paragraph*{Contributions.}
	
	HZ, JK, and RM discovered the phenomenon. 
	LJIM, PMS, MB, and AA conceived the experimental realization.
	PMS, YS, and HZ worked out the theory details.
	LJIM and ED implemented and built the experiment.
	LJIM collected experimental data.
	LJIM and PMS analyzed experimental data.
	PMS and YS did simulations.
	MB and AA supervised the theory and experimental work. 
	All authors contributed to the manuscript.

	%%%%%%%%%%%%%%%%%%%%%%%%%%%%%%%%%%%%%%%%%%%%%%%%%%%%%%%%%%%%%%%%%%%%%%%%%%%
	%                        Methods
	%%%%%%%%%%%%%%%%%%%%%%%%%%%%%%%%%%%%%%%%%%%%%%%%%%%%%%%%%%%%%%%%%%%%%%%%%%%
	%%%%%%%%%%%%%%%%%%%%%%%%%%%%%%%%%%%%%%%%%%%%%%%%%%%%%%%%%%%%%%%%%%%%%%%%%%%
	
	%\newpage
	\beginmethods
	
	\section*{Methods}
	
	\subsection*{Setup}
	Experiments here employed a single crystal diamond with $\thicksim$ 1ppm NV centres and natural abundance ($1.1\%$) of \C\ nuclei. The diamond sample, immersed under water, is mounted in a 8mm 7inch glass sample tube, such that the [100] face is parallel to external magnetic field with magnitude $B_{0}$. The tube is, in turn, attached to a carbon-fiber rod, using two O-rings, and mounted the rod to a belt-drive actuator (Parker) that ``shuttles” that sample rapidly between fields used for \C\ hyperpolarization ($B_{0}{=}38$mT) and \C\ interrogation ($B_{0}{=}7$T). 
	
	The \C\ nuclei are hyperpolarized for $t_{\R{pol}}{=}60$s via NV centers at low magnetic field (38mT) via a CW laser illumination and chirped microwave protocol described in Ref.~\cite{ajoy2018orientation}, and following a spin-ratchet polarization transfer mechanism described previously in  Refs.~\cite{zangara2019dynamics,pillai2023electron}. The hyperpolarization setup uses multi-laser excitation and has been described previously~\cite{sarkar2022rapidly}. At high field ($B_{0}{=}7$T), the \C\ nuclei are subsequently subjected to the random multipolar drives, with the \C\ Larmor precession being sampled in windows between the spin-locking sequences (Fig.~\ref{fig:intro}c).
	
	\subsection*{Spin control architecture}
	A particular innovation in the current experiments is the design of a new spin control infrastructure that facilitates versatile spin control. Hundreds of thousands of pulses are typically applied, and while in previous experiments~\cite{DTC_Beatrez2022}, the pulses needed to be all identical due to memory and other technical limitations, here we significantly lift this constraint. We accomplish this by constructing a new NMR spectrometer fully based on a high-speed large-memory Arbitrary Wave Generator (AWG) (Tabor P9484M). The AWG is employed to construct the RF pulses, which are then amplified by a Traveling-Wave Tube (Herley) amplifier and delivered to the RF coil via a cross-diode based transmit/receive (T/R) transcoupler (Tecmag). The rapid sampling rate (up to 9GS/s) and substantial onboard memory (16GB) of the AWG, along with the ability to use onboard Numerically Controlled Oscillators (NCOs), provides a versatile control toolbox; in principle, any sequence of RF pulses can be applied to the spins, comprising over ${>}64000$ unique building blocks, and any larger combination thereof. 
	
	In addition, the AWG is also used as a fast digitizer to rapidly sample (here at 1GS/s) the \C\ Larmor precession between the pulses. The inductively measured signal is amplified via a chain of low-noise amplifiers (ARR and Pasternack) through the T/R transcoupler prior to digitization. Using the same onboard NCOs can now down-shift the precession signals to obtain in-phase and quadrature components. This allows us to interrogate the x-projection and y-projection of the \C\ nuclear spins directly in their rotating frame.
	
	To create the \textit{rondeau} order using this new capability, we first create two waveforms, which correspond to\pluspole and\minuspole.\pluspole and\minuspole are defined as: $N_\mp$ $\frac{\pi}{2}_x$ pulses, $y$-pulse of angle $\gamma_y$, and $N_\pm$ $\frac{\pi}{2}_x$ pulses, with free evolution time $t_{\text{free}}$ after every pulse ($N_{+} {=} 200, N_{-} {=} 100$). We then generated the sequence by placing the two waveforms,\pluspole and\minuspole, in the desired order of application. To read out the signal of the \C\ nuclear spins after each pulse, we first wait for $t_{\text{ring-down}}{\approx} 12\mu\mathrm{s}$ to account for any pulse ring-down, followed by inductive detection for $t_{\text{acq}} {=} 12\mu\mathrm{s}$. Thus, in total, the spacing in between the pulses  
	is $t_{\text{free}} = t_{\text{ring-down}} +  t_{\text{acq}}$. 
	
	\subsection*{Random multipolar drives}
	
	Temporal order is realized using the `slow' drive,  which follows a structured RMD sequence.
	More precisely, the protocol includes two elementary building blocks $U_0^{\pm}$ of equal duration $T$:
	\begin{equation}
		U_0^{\pm} = (U_x U_\mathrm{dd})^{N-N_\pm} U_y U_\mathrm{dd} (U_x U_\mathrm{dd})^{N_\pm}\,,
		\label{eq:U_plusminus}
	\end{equation}
	where $U_x=\exp(-i\theta_x I_x)$, $U_y=\exp(-i\gamma_y I_y)$, $U_\mathrm{dd}=\exp(-i\tau H_\mathrm{dd})$. We neglect the dipole-dipole interaction when $x,y$-pulses are applied, as the Rabi frequency $\Omega$ is much larger than the median coupling $J$, $\Omega{>}10J$. In the main text, we use \pluspole/\minuspole to denote $U_0^+$/$U_0^-$ for simplicity. $U_xU_{\mathrm{dd}}$ implements a spin-lock pulse and a free evolution governed by $H_{\mathrm{dd}}$, and $U_y$ implements the polarization inversion. Trains of spin-lock cycles generate the emergent $U(1)$ symmetry required for the quasi-conservation of the polarization. Note that different numbers of spin-lock pulses are deployed before $U_y$ in these two trains. Hence, polarization flips at different times if different blocks are applied to the system. 
	
	For Floquet protocols, the system propagates deterministically with, for instance, only $U_0^+$. In contrast, for $n{=}0$ random multipolar driving, the two operators (or monopoles) are randomly selected to evolve our system. Since this selection is random in time, its DFT is trivially flat. Note that the specific construction of $U_0^{\pm}$ already embeds a certain structure in the protocol, for instance, polarization flips precisely once within a period. For comparison, in completely structureless random drives, polarization flip may happen at any time, which normally melts the long-range temporal order rapidly.  
	
	Higher order multipolar operators of order $n$ can be recursively constructed as $U_n^{\pm}{=}U_{n-1}^{\mp}U_{n-1}^{\pm}$, by anti-aligning $n{-}1$-multipole pairs together. The length of an $n$-multipole sequence grows exponentially in $n$ as $2^nT$.
	In complete analogy, the $n-$RMD protocol consists of a sequential application of a random selection of $U_n^{\pm}$ with equal probability. In the limit $n{\to}\infty$, the protocol becomes deterministic and quasiperiodic in time. It indeed corresponds to the Thue-Morse sequence, which has also been extensively studied in the context of, for instance, quasi-crystals and number theory~\cite{avishai1992transmission,allouche1999ubiquitous}.
	
	\subsection*{Prethermal order}
	
	Generic time-dependent many-body systems do not obey the energy conservation law. Therefore, they tend to absorb energy from the external drive, and eventually heat up towards a featureless state at infinite temperature. In Floquet systems, heating can be significantly suppressed if there is a notable mismatch between the local energy scale and the external driving frequency, for instance, in the high-frequency regime~\cite{ho2023quantum}. This can result in an exceptionally long-lived prethermal regime before notable heating happens~\cite{Abanin2015_Heating,Mori2016_Heating,rubio2020floquet,peng2021floquet}, and dynamics therein can be approximated by a static quasi-conserved effective Hamiltonian $H_{\mathrm{eff}}$. It can be perturbatively constructed, for instance, by using a Floquet Magnus expansion or high-frequency expansion~\cite{goldman2014periodically,eckardt2015high}. 
	
	For ergodic interacting many-body systems, the existence of the effective Hamiltonian implies that, during the prethermal regime, local properties of the system can be captured by a prethermal canonical ensemble $\rho_{\mathrm{pre}}{\sim} e^{-\beta_{\mathrm{eff}} H_{\mathrm{eff}}}$; here $\beta_{\mathrm{eff}}$ denotes the prethermal temperature that is determined by the energy density of the initial state~\cite{Mori2016_Heating}. If $\beta_{\mathrm{eff}}$ is sufficiently low and $H_{\mathrm{eff}}$ allows for spontaneous symmetry breaking to occur at a finite temperature, $\rho_{\mathrm{pre}}$ can exhibit equilibrium spatial ordering. Additionally, if regular polarization inversion is further introduced by the drive, as described in the last section, prethermal non-equilibrium time crystalline order can form~\cite{machado2020long,pizzi2021classical}. 
	
	This prethermal phenomenon can be generalized to other time-dependent protocols even in the absence of strict temporal periodicity. One typical example is quasiperiodically driven systems where at least two drive frequencies are incommensurate with each other~\cite{martin2017topological,dumitrescu2018logarithmically,zhao2019floquet,crowley2019topological,else2020long,wen2021periodically,nathan2022topological,he2024experimental,chen2024multifractality}. Prethermalization also occurs in RMD systems, where the driving spectrum has continuous support over the entire range of frequencies due to temporal randomness~\cite{RMD_Intro_Zhao_etal2021}. 
	
	The temporal multipolar correlation of RMD protocols significantly modifies the Fourier spectrum of the random drive sequence: It has been shown in Ref.~\cite{RMD_Intro_Zhao_etal2021} that the envelope of the spectrum follows
	$\prod_{j=1}^n [1-\cos \left(2^{j-1} \nu\right)]^{1/2}$ with $\nu$ the Fourier frequency. This expression indicates an algebraic suppression $\nu^n$ for $\nu{\to} 0$. Since it is these low-frequency modes that normally produce the dominant contribution to heating, heating can be algebraically suppressed in the high-frequency regime. The algebraic scaling exponent also has an explicit dependence on the multipolar order. One can also construct effective Hamiltonians by generalizing the Floquet-Magnus expansion, and using linear response theory to analyse the heating behavior more systematically~\cite{RMD_Rigorous_Mori_etal2021}.
	
	If the temporal randomness only weakly perturbs the system, the RMD protocol indeed leads to
	a prethermal plateau, that can feature either equilibrium or non-equilibrium ordering, very similar to Flqouet systems. However, the RMD protocol introduced in the last section is designed such that temporal randomness in polarization inversion strongly changes the behavior of the micromotion of the system. As illustrated in the main text, this RMD protocol results in rondeau order, beyond the conventional Floquet paradigm.
	
	It is worth noting that, although those prethermal orders eventually melt, their stability and lifetime can be parametrically controlled, permitting direct experimental observation with our current nuclear spin setup. Throughout, we estimate the lifetime via the $1/e$ decay time $T_e$ defined as the time where the absolute value of the signal is closest to $1/e$ of its initial value $S_0$, i.e.,
	\begin{equation}\label{eq:lifetime}
		T_e = \underset{t>0}{\mathrm{argmin}}\, \abs{ \abs{S(t)} - S_0/e }\, . 
	\end{equation}
	This feature may also be exploited to study interesting applications [see main text]. 
	
	\subsection*{Discrete Fourier Transform}
	
	In the main text, we use two different discrete Fourier transforms. For the micromotion we consider the discrete Fourier transform of the digitized micromotion extracted at times $(2\ell+1) T/2$ ($\ell{\in}\mathbb{N}$):
	\begin{equation}\label{eq:micro_DFT}
		\mathrm{DFT}_\mathrm{micro}(\omega_k) 
		= \frac{1}{M} 
		\sum_{\ell=0}^{M-1} e^{-i\omega_k \ell} \mathrm{sgn}\left(S((2\ell+1) T/2)\right) \, ,
	\end{equation}
	where $\omega_k = k 2\pi/M$, $k=0,\,\dots, M-1$ and $\mathrm{sgn}\left(S(t)\right)=+1$ or $-1$ depending on whether $S(t)>0$ or $S(t)<0$, respectively.
	
	For the stroboscopic evolution we consider the discrete Fourier transform of the signal extracted at stroboscopic times $\ell T$~($\ell{\in}\mathbb{N}$):
	\begin{equation}\label{eq:strobo_DFT}
		\mathrm{DFT}_\mathrm{strobo}(\omega_k) 
		= \frac{1}{M} 
		\sum_{\ell=0}^{M-1} e^{-i\omega_k \ell} S(\ell T) \, .
	\end{equation}
	Note, that the Fourier amplitude is defined as the absolute value of the DFT, $(\mathrm{Fourier\,amplitude}{=}\abs{\mathrm{DFT}})$, in contrast to the Fourier intensity (i.e., the power spectrum) which is defined as the absolute value squared.

	%%%%%%%%%%%%%%%%%%%%%%%%%%%%%%%%%%%%%%%%%%%%%%%%%%%%%%%%%%%%%%%%%%%%%%%%%%%
	%                    Bibliography
	%%%%%%%%%%%%%%%%%%%%%%%%%%%%%%%%%%%%%%%%%%%%%%%%%%%%%%%%%%%%%%%%%%%%%%%%%%%
	%%%%%%%%%%%%%%%%%%%%%%%%%%%%%%%%%%%%%%%%%%%%%%%%%%%%%%%%%%%%%%%%%%%%%%%%%%%
	
	%\clearpage
	\bibliography{biblio_rmd}
	\vspace{-1mm}

	%%%%%%%%%%%%%%%%%%%%%%%%%%%%%%%%%%%%%%%%%%%%%%%%%%%%%%%%%%%%%%%%%%%%%%%%%%%
	%                        SI
	%%%%%%%%%%%%%%%%%%%%%%%%%%%%%%%%%%%%%%%%%%%%%%%%%%%%%%%%%%%%%%%%%%%%%%%%%%%
	%%%%%%%%%%%%%%%%%%%%%%%%%%%%%%%%%%%%%%%%%%%%%%%%%%%%%%%%%%%%%%%%%%%%%%%%%%%

	\clearpage
	\onecolumngrid
	%\begin{widetext}
	\begin{center}
		\textbf{\large{\textit{Supplemental Information:} \\ \smallskip Experimental observation of a time rondeau crystal:\\ Temporal Disorder in Spatiotemporal Order}} \\\smallskip
	\end{center}

	\twocolumngrid
	
	\beginsupplement
	\tableofcontents

	%%%%%%%%%%%%%%%%%%%%%%%%%%%%%%%
	%            Text
	%%%%%%%%%%%%%%%%%%%%%%%%%%%%%%%
	%%%%%%%%%%%%%%%%%%%%%%%%%%%%%%%

	\section{Experiment}

	\subsection{Inhomogeneity of pulses}

\begin{figure}[t]
		\centering
		\includegraphics[width=0.4\textwidth]{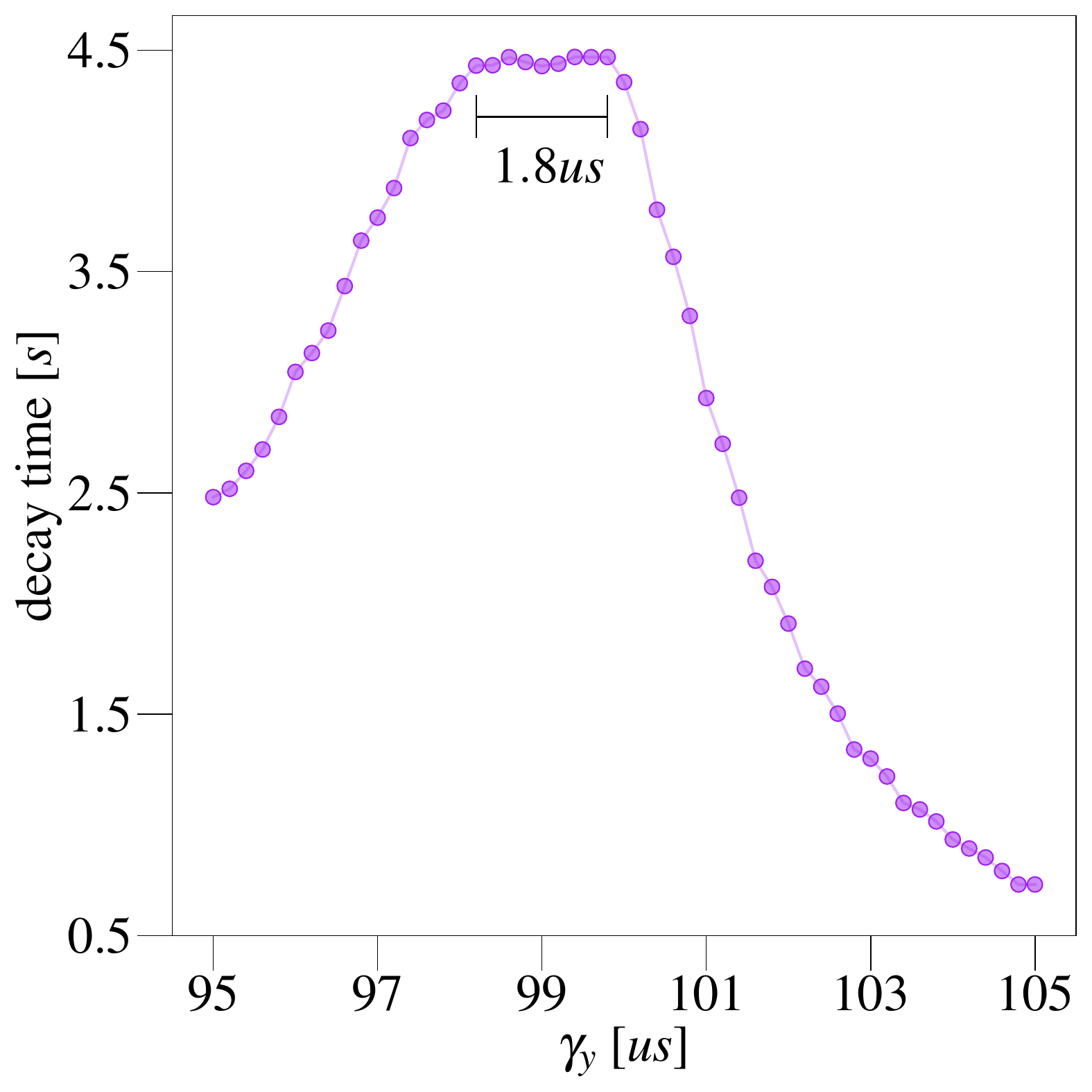}
		\caption{\textbf{Change in the decay time of the DTC while varying length of $\mathbf{\gamma_{y}}$ (experiment).} Change in the decay time of the DTC while varying the length of $\gamma_{y}$ near $\pi$. Plateau of length $1.8\,\mu\mathrm{s}$ indicates the inhomogeneity of the $\gamma_{y}$ pulse applied to the sample.} 
		\label{sup_fig:rf_inhomo}
	\end{figure}

	Due to the finite-size of the sample and the restriction in the coil design, the RF coil in our apparatus (described in Supplementary Figure 2 of \cite{sahin2022high}) is incapable of applying perfectly homogeneous transverse magnetic field to the sample. To quantify the inhomogeneity, we show the change in the decay rate of the discrete time crystal (DTC) order engineered using two frequency drive described in \cite{DTC_Beatrez2022} as we incrementally change the length of the $\gamma_{y}$ pulse. Fig.~\ref{sup_fig:rf_inhomo} shows the region of length $1.8\,\mu\mathrm{s}$ where the decay of the DTC remains approximately constant. This region of plateau indicates that the inhomgeneity of the pulses is $\approx 1.8\,\mu\mathrm{s}/100\,\mu\mathrm{s} = 1.8\%$, because the \C\ nuclear spins in the same sample feels different magnetic fields depending on their positions. Here, $100\,\mu\mathrm{s}$ is used to approximately deduce the length of the $\pi\text{-pulse}$.

	\subsection{Maximum length of encoding}
	
\begin{figure}[t]
		\centering
		\includegraphics[width=0.45\textwidth]{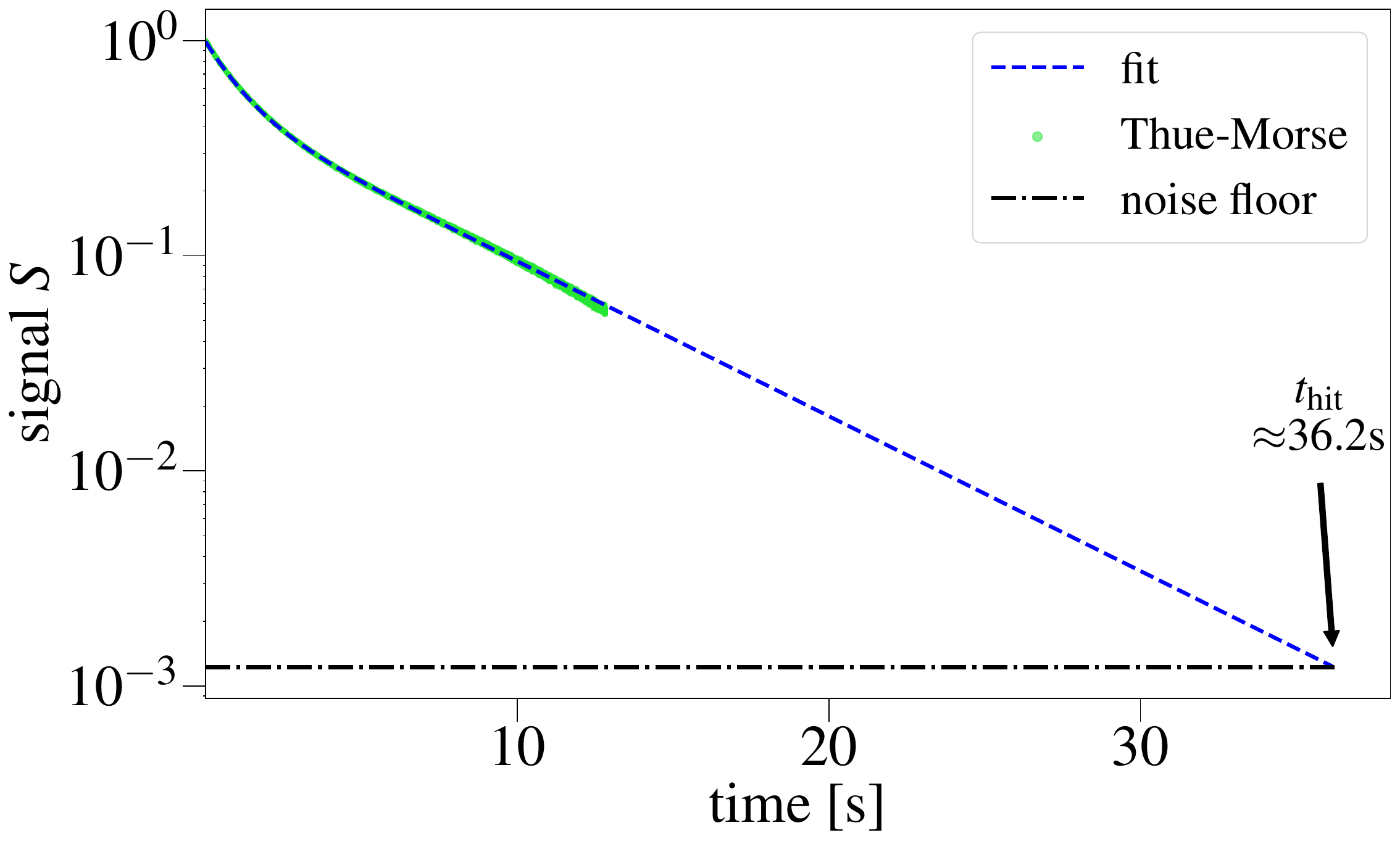}
		\caption{\textbf{Estimating the maximum length of encoding.} Green points indicate normalized signal of the \C\ nuclear spins when Thue-Morse sequence described in the main text Fig.~\ref{fig:intro}a ($n \rightarrow \infty$) is applied. Dotted blue line fits the signal of \C\ nuclear spins to a biexponential function. From the fit, the signal is estimated to hit the noise floor (black dotted line) at $t_{\text{hit}} \approx 36.2\R{s} \approx 1.3\times10^{3}T$. We can thus encode up to 198 characters, using 7-bit ASCII encoding scheme.
		}
		\label{sup_fig:encoding}
	\end{figure}

	To obtain a conservative estimate of the maximum-length encoding that can be obtained in our experimental setup, we first fit the decay of the \C\ nuclear spins' signal generated by Thue-Morse sequence (described in Fig.~\ref{fig:intro}a as $n \rightarrow \infty$) to a biexponential function, as we empirically observed that the biexponential function (dotted blue line in Fig.~\ref{sup_fig:encoding}) fit the raw data (green points in Fig.~\ref{sup_fig:encoding}) well with high fidelity. We further assume that the decay time of any string-encoded time rondeau crystal is comparable to that of the Thue-Morse sequence, as shown for one example in main text Fig.~\ref{fig:info_enc}.
	
	Fig.~\ref{sup_fig:encoding} shows that at $t_{\text{hit}} \approx 36.2s$, the signal is estimated to hit the noise-floor. With $t_{\text{hit}}$ as our bandwidth, we can encode up to $198$ characters, which corresponds to approximately $1.3\times 10^{3}$ number of cycles. For future applications, we can improve the bandwidth by increasing the initial signal with better hyperpolarization of the \C\ nuclear spins. We can further improve the bandwidth by increasing the Rabi-frequency using a RF coil with higher Q-factor and filling factor, as short, high power pulses decreases the decay caused by dipole-dipole coupling during the pulses.
	
	\section{Simulations}
	
	\subsection{Model}
	
	The experiment is performed in a strong magnetic field~(${\sim}7\,\mathrm{T}$). The Hamiltonian reported on in the main text, Eq.~\eqref{eq:dip}, describe the relevant dipole-dipole interaction in a rotating frame that removes this strong magnetic field.
	In this section, we briefly report on the derivation of this rotating frame.
	
	The lab frame Hamiltonian of the interacting system of \C-nuclear spins is given by,
	\begin{equation}\label{SI:eq:lab_hamiltonian}
		\mathcal{H}_\mathrm{lab}\lb t\rb = \omega_L I_z + H_{\mathrm{dd},\mathrm{lab}} + H_{x,\mathrm{lab}}(t) + H_{y,\mathrm{lab}}(t) \, ,
	\end{equation}
	where $\omega_L$ is the Larmor frequency, and
	\begin{equation*}
		H_\mathrm{dd}=\sum_{k<l} J_{kl} \left[3 r^{-2}_{kl}(\boldsymbol{I}_k \cdot \boldsymbol{r}_{kl})(\boldsymbol{I}_l \cdot \boldsymbol{r}_{kl})  - \boldsymbol{I}_k \cdot \boldsymbol{I}_l\right]\, ,
	\end{equation*}
	is the magnetic dipole-dipole interaction with the vector $\boldsymbol{r}_{kl}$ connecting the two spins $k$ and $l$. The time-dependent terms in Eq.~\eqref{SI:eq:lab_hamiltonian} are given by $H_{x,\mathrm{lab}}(t)=\Omega \cos(\omega_L t) \Theta_x(t) I_x$ and $H_{y,\mathrm{lab}}(t)=\Omega \cos(\omega_L t) \Theta_y(t) I_y$, with $\Theta_{x,y}(t)$ being step functions that implement the sequence described in Fig.~\ref{fig:intro}.
	Performing a co-moving frame transformation with respect to the magnetic field and a rotating wave approximation we arrive at
	\begin{equation}\label{SI:eq:rot_hamiltonian}
		\mathcal{H}_\mathrm{lab}\lb t\rb = H_{\mathrm{dd}} + \Theta_x(t) \Omega I_x + \Theta_y(t) \Omega I_y \, ,
	\end{equation}
	with the secular dipole-dipole interaction
	\begin{equation*}
		H_{\mathrm{dd}}=\sum_{k<l} J_{kl} \left[3 I_k^z I_l^z  - \boldsymbol{I}_k \cdot \boldsymbol{I}_l\right]\, ,
	\end{equation*}
	where all terms that do not commute with the $I_z$ field are cancelled, thus arriving at the Hamiltonian reported in Eq.~\eqref{eq:dip}.
	
	Note that in general, the interaction term~($H_{\mathrm{dd}}$) is also present during the application of the $x$ and $y$ pulses. However, since the Rabi-frequency~($\Omega$) is much larger than the median coupling~($J\ll\Omega$), we ignore the impact of interactions when the pulses are applied for numerical simplicity. This results in the unitary evolution Eq.~\eqref{eq:U_plusminus}.

	\begin{figure}[t]
		\centering
		\includegraphics[width=0.5\textwidth]{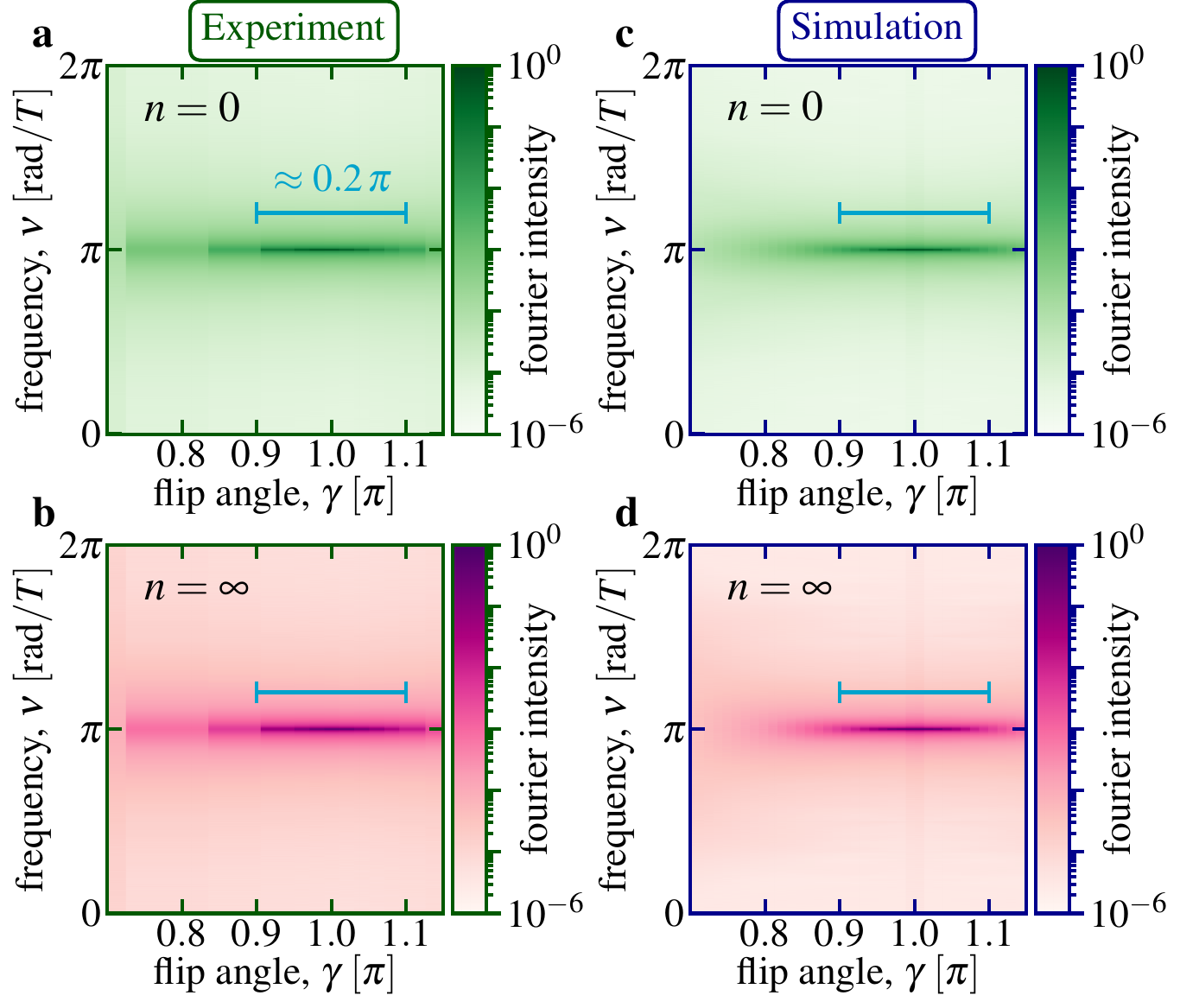}
		\caption{
			\textbf{Comparison of experimental and simulated phase diagram.} 
			Intensity of Fourier transform of stroboscopic dynamics for monopole~($n=0$), \textbf{a}, and Thue-Morse sequence~($n=\infty$), \textbf{b}, similar to main text Fig.~\ref{fig:phase_diagram}; and corresponding simulations in \textbf{c} and \textbf{d}, respectively.
			The simulation results align perfectly with the experimental observation.
			The strong response of the Fourier intensity at $\nu=\pi/T$ is a clear signature for period doubling dynamics that are stable in a large parameter regime, $\gamma_y \approx \pi \pm 0.1\pi$. The stability seems to be independent of the multiple order~($n$). 
			Parameters are as in main text Fig.~\ref{fig:phase_diagram}.
		}
		\label{SI:fig:phase_diagram_comparison}
	\end{figure}
	
	\subsection{Algorithm}

	The algorithm we use to numerically simulate the system is similar to that used in Ref.~\cite{DTC_Beatrez2022}.
	Spins are placed on a pseudo-random graph (among the majority of $^{12}$C spins which carry no total nuclear spin, the position on the diamond lattice of \C\ spins with total spin-$1/2$ is random), coupled with long-range dipole-dipole interactions, see Eq.~\eqref{eq:dip}. The pseudo-random graph is generated with the procedure designed in Ref.~\cite{DTC_Beatrez2022}: Spin locations are drawn one-by-one randomly in a $3$D cube with edge length $d$. Note that the dipole-dipole coupling strongly depends on the distance of the spins, $J_{k\ell}\propto r_{k\ell}^3$. Thus, if two spins are placed close to one another they will interact strongly effectively decoupling from their environment; to avoid this scenario a spin-location is only accepted if the distance to all other spins is at least $r_{\text{min}}$. Likewise, if a spin is located far a way from all other spins, it also effectively decouples from all other spins. To make maximal use of the given number of spins, we ensure that each spin has a distance of at most $r_{\text{max}}$ to at least one of the other spins.
	We choose $r_{\text{min}}=0.9$ and $r_{\text{max}}=1.1$.
	
	In the simulations, unless otherwise noted, we initialize all spins in the all $\hat{x}$-polarized pure state. We perform time evolution according to the protocols described by Fig.~\eqref{fig:intro} and Eq.~\eqref{eq:U_plusminus} using exact-diagonalization provided by the QuSpin python library~\cite{quspin2017,quspin2019}. As mentioned above, $x$ and $y$ pulses can be treated as instantaneous pulses for numerical efficiency.
	Although not shown here, we also have performed simulations with finite-time pulses
	and indeed, found no notable difference in the observed dynamics.

	\begin{figure}[t]
		\centering
		\includegraphics[width=0.5\textwidth]{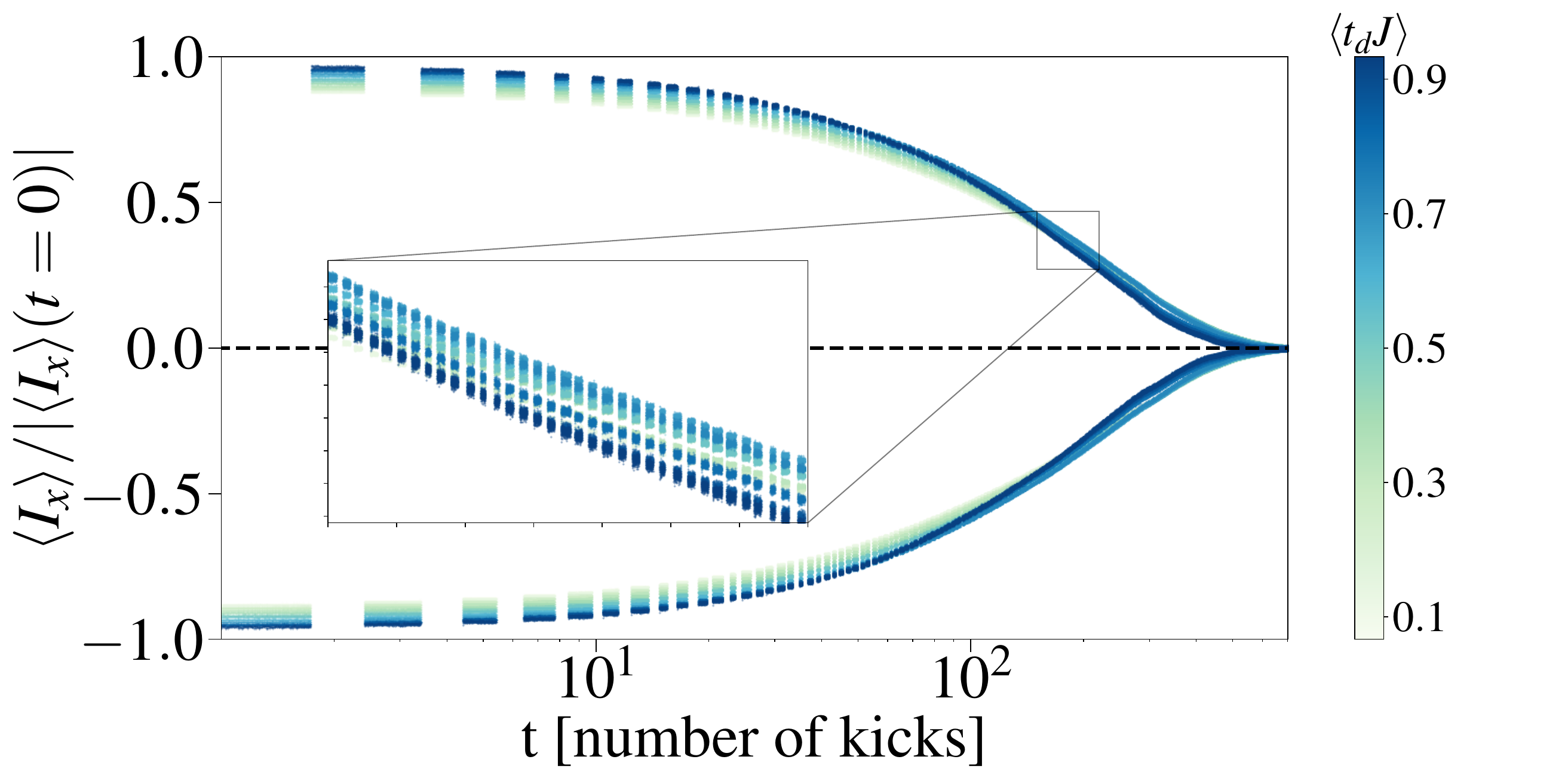}
		\caption{
			\textbf{Lack of initial state dependence in time rondeau crystal.}
			Normalized $I_x$ polarization as a function of time for different delay times $t_d$. 
			The initial state is subject to free-induction-decay caused by applying the dipole-dipole interactions~($H_\mathrm{dd}$, Eq.~\eqref{eq:dip}) for a time $t_d$, indicated by colorbar.
			The time evolution is independent of the initial decay time $t_d$, indicating that the formation of temporal order is independent of the initial state.
			We use $\gamma_y=\pi$; other parameters are as in Fig.~\ref{fig:intro}
		}
		\label{sup_fig:init_state}
	\end{figure}

	\subsection{Comparison of Experiment and Simulations}
	
\begin{figure}[t]
		\centering
		\includegraphics[width=0.5\textwidth]{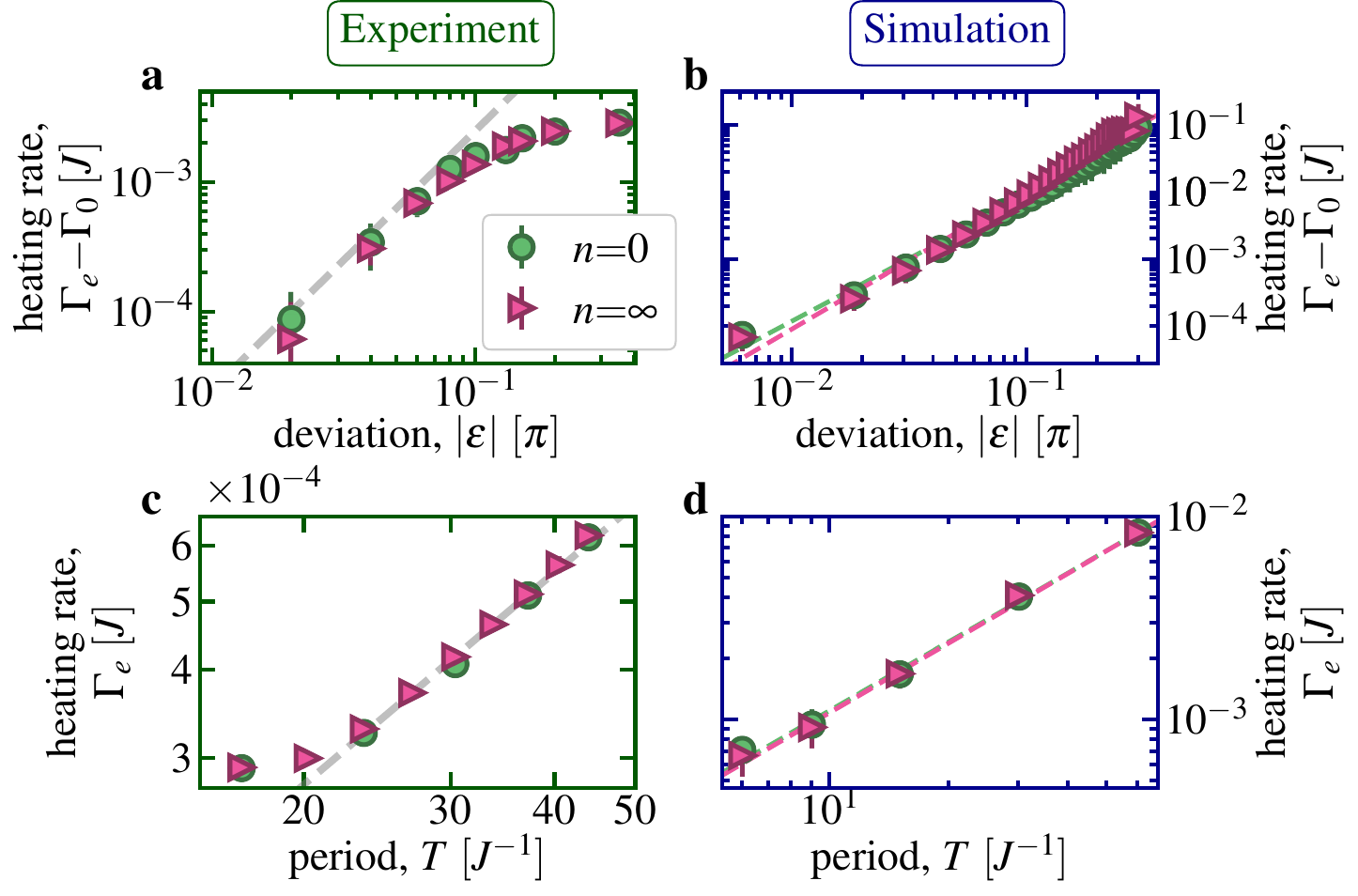}
		\caption{
			\textbf{Experimental and simulated lifetime dependence.}
			\textbf{a}, heating rate dependence on the deviation ($\varepsilon = \gamma_y -\pi$), for monopole~($n=0$, green circles) and Thue-Morse~($n=\infty$, pink triangles); gray lines indicate power law $\varepsilon^2$. Data is same as displayed in Fig.~\ref{fig:heating_rate}a
			.   \textbf{b}, corresponding numerical simulation; dashed lines indicate power law fits with exponent $1.9\pm0.2$ and $2.0\pm0.2$ for monopole and Thue-Morse sequence, respectively.
			\textbf{c}, dependence of heating rate on period $T$, while simultaneously varying $\varepsilon=BT$; gray lines indicate power law $T^1$. Data is same as displayed in Fig.~\ref{fig:heating_rate}b.
			\textbf{d}, corresponding numerical simulation; dashed lines indicate power law fits with exponents $1.1\pm0.1$ for both monopole and Thue-Morse sequence.
			The simulation results are in good quantitative agreement with the experimental results. The extracted power laws are consistent with the dephasing limit predictions [see text], $\Gamma_e-\Gamma_0 \propto \varepsilon^2$ and $\Gamma_e \propto T$ for the scaling with small detuning~($\varepsilon$) and period~($T$), respectively.
			Parameters are as in main text Fig.~\ref{fig:heating_rate}.
		}
		
		\label{SI:fig:lifetime_comparison}
	\end{figure}

	Equipped with this algorithm we can now support the experimental findings with corresponding numerical simulations. 
	Let us emphasize that, in the experiment, the diamond sample normally contains hundreds of NV-centers each surrounded by thousands of spins, and hence the precise numerical simulation of its dynamics is far beyond the reach of classical computational power.
	Yet, we find good qualitative agreement between simulations with $14$ spins and experiments for both the phase diagram and decay rate scaling:
	In Fig.~\ref{SI:fig:phase_diagram_all}, we compare the experimentally obtained phase diagram~(cf.~Fig.~\ref{fig:phase_diagram}) with simulation results. The simulations align perfectly well with the experimental measurements, indicating a similar regime of stability and a lack of dependence on the stability of the drive sequence.
	In addition, in Fig.~\ref{SI:fig:lifetime_comparison}, we present results for the heating rate $\Gamma_e$. Again, we find good qualitative agreement. Specifically, both simulation and experiment are consistent with a power law of the heating rate $\Gamma_e-\Gamma_0\propto\varepsilon^2$ and $\Gamma_e \propto T$ for small $\varepsilon$ and $T$, see Fig.~\ref{SI:fig:lifetime_comparison}a, b and c, d, respectively.
	
	\begin{figure}[t]
		\centering
		\includegraphics[width=0.5\textwidth]{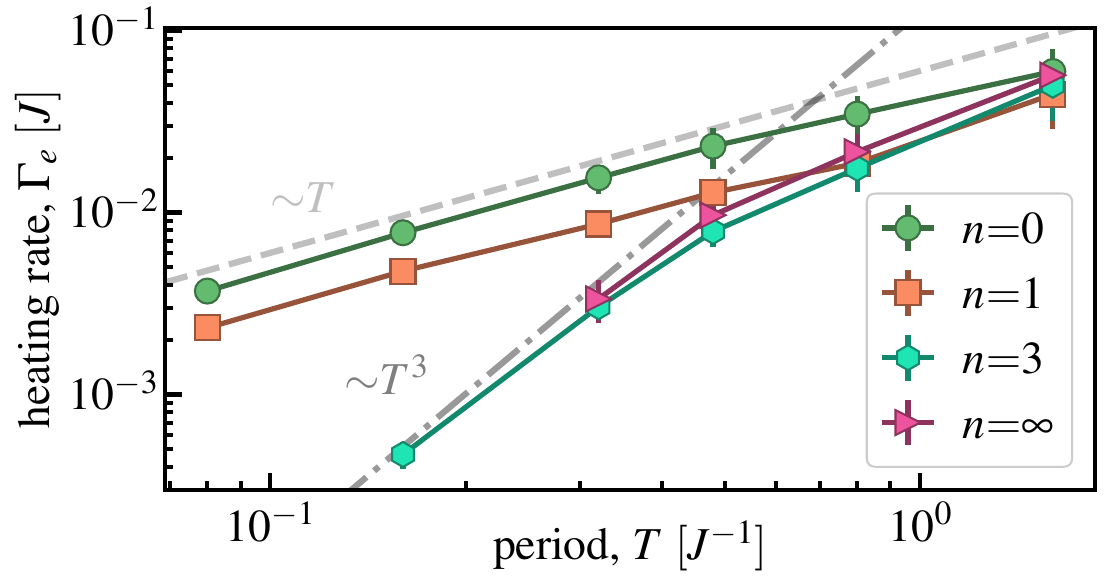}
		\caption{
			\textbf{Simulated heating rates, approaching the high-frequency ($T\to 0$) regime.}
			Heating rate for $n=0$~(green circles), $n=1$~(orange squares), $n=3$~(cyan hexagons) and Thue-Morse~($n=\infty$, pink triangles) obtained from numerical simulations with $(N_+, N_-)=(10,5)$. Gray dashed and dash-dotted lines are guides to the eye, corresponding to power laws $\sim T$ and $\sim T^3$, respectively.
			We observe a dependence of the heating rate on the multipole order $n$, unlike in the experiments, and corresponding simulations, shown in the main text~(see also Fig.~\ref{SI:fig:lifetime_comparison}). This indicates that a dependence of the heating rate on the drive protocol occurs in the high-frequency limit. Note that the period values here are at least one order of magnitude smaller than the experimentally accessible regime, see Fig.~\ref{fig:heating_rate}.
			Other parameters are as in Fig.~\ref{SI:fig:lifetime_comparison}.
		}
		\label{SI:fig:high_frequency}
	\end{figure}

	\subsection{Dephasing limit and low drive frequency regime}
	
	As outlined in the main text, our experimental results and the corresponding numerical simulations of the heating rate do not match with earlier theoretical predictions~\cite{RMDDTC_Zhao_etal2023}. In Ref.~\cite{RMDDTC_Zhao_etal2023}, a power law suppression of heating rate $\Gamma_e \propto T^{\beta}$ is predicted in the high-frequency regime, $JT{\ll} 1$. The exponent $\beta$ increases notably with the multipole order $n$. This happens because of the stronger low-frequency suppression of the driving spectrum at higher $n$, as confirmed in the DFT spectrum, cf.~Fig.~\ref{fig:Floquet_vs_RMD}. 
	However, using a `slow drive', our experiments are performed in a low- to intermediate-frequency regime~($JT{>}1$). Therefore, we find little to no dependence on the multipole order $n$ in our experiments, see Fig.~\ref{fig:heating_rate}.
	
	The parametric dependence of the lifetime (inverse heating rate) versus $T$ and $\varepsilon$ can be understood in a dephasing limit as follows.
	The imperfection in the $y$-pulses, $\varepsilon \neq 0$, results in an imperfect polarization inversion $I^x \to - \cos(\varepsilon) I^x + \sin(\varepsilon) I^z$. Only the $\xhat$-polarization can be prethermnally protected by the fast drive, and other components are rapidly echoed out between two $y$-pulses. Therefore, the signal after one full cycle is given by $I^x \to - \cos(\varepsilon) I^x$, leading to the period-doubling oscillations at stroboscopic times with an overall weakly decaying amplitude. Factoring in the intrinsic heating rate~($\Gamma_0$) due to imperfect quasi-conservation of $I^x$,  the amplitude after $M$ periods can be approximated as $\abs{I^x}\propto \cos(\varepsilon)^M e^{-\Gamma_0 M T}$.
	The independence of the heating rate on the multipolar order $n$ follows immediately: the above heating only depends on the number $M$ of $y$-pulses applied within one period $T$, which is the same for all sequences.
	
	Considering small detuning values~($\varepsilon\ll 1$), we can expand the above equation as $\abs{I^x}\propto (1-\varepsilon^2/2)^M e^{-\Gamma_0 M T}\approx e^{-M\varepsilon^2/2-\Gamma_0 M T}$. Upon recasting this equation as $\abs{I^x}\propto e^{-\Gamma_e M T}$, one finds a power law scaling for the heating rate, $\Gamma_e-\Gamma_0 \propto \epsilon^2$. The numerical simulations are consistent with this exponent, see Fig.~\ref{SI:fig:lifetime_comparison}b.
	
	In addition, if $\Gamma_0$ is sufficiently small compared to the above heating rate due to the inversion imperfection, and using the dependence  $\varepsilon= B T $, one finds $\Gamma_e \approx \Gamma_e - \Gamma_0 \propto  T$. The numerical simulations are consistent with this linear dependence, see Fig.~\ref{SI:fig:lifetime_comparison}d.
	
	In fact, also the bending-up of the experimental curve in the regime $T\to 0$ that is observed in the experimental data shown in Fig.~\ref{SI:fig:lifetime_comparison}c, can be understood as follows. First, let us reiterate on the fact that in the experiment there is a finite uncertainty in determining the pulse-angle $\gamma=\pi$, see Fig.~\ref{sup_fig:rf_inhomo}. Thus, we should rather consider $\varepsilon =\Delta \varepsilon + B T $ where $\Delta \varepsilon$ is the uncertainty in $\varepsilon$. This modifies $\Gamma_e - \Gamma_0 \propto (\Delta \varepsilon)^2/T + 2B\Delta \varepsilon + B^2 T$ which reaches a minimum where it flattens. Eventually, we expect the rate to start increasing for sufficiently small $T$; however, let us stress, that in general $\Gamma_0$ also depends on the period $T$, since it controls the spin-locking lifetime.
	
	Finally, let us point out that the starting point for this analysis -- the dephasing limit -- is expected to break down in the high-frequency regime, where we expect that the decay rate explicitly depends on the multipolar order
	~\cite{RMDDTC_Zhao_etal2023}. 
	However, this high-frequency regime is currently inaccessible in our experiments due to Rabi frequency limits on the \C\ nuclei, which puts a lower limit on the pulse-duration needed to achieve the $(\pi/2)_x$ and $\gamma_y$ pulses; thus, putting a lower limit on the period $T$ that is experimentally achievable. Note, however, that this is not a fundamental constraint and can be overcome by employing higher Q-factor RF coils and increasing the sample filling factor (currently ${<}$10\%).
	While exploring this regime is beyond the reach of the present experiment, we can still perform numerical simulations to demonstrate the occurrence of this phenomenon. In Fig.~\ref{SI:fig:high_frequency}, for a small driving period, the heating rate can be different for various multipole order~($n$).
	Interestingly, the power law scaling exponents shown in Fig.~\ref{SI:fig:high_frequency} do not follow the prediction in Ref.~\cite{RMDDTC_Zhao_etal2023}, where the exponent $2n+1$ is predicted for locally interacting many-body systems. We attribute this discrepancy to the presence of long-range interactions in the experimental system (in contrast to the short-range model studied in Ref.~\cite{RMDDTC_Zhao_etal2023}), but we leave a proper analysis to future work.
	
	\begin{figure}[t]
		\centering
		\includegraphics[width=0.5\textwidth]{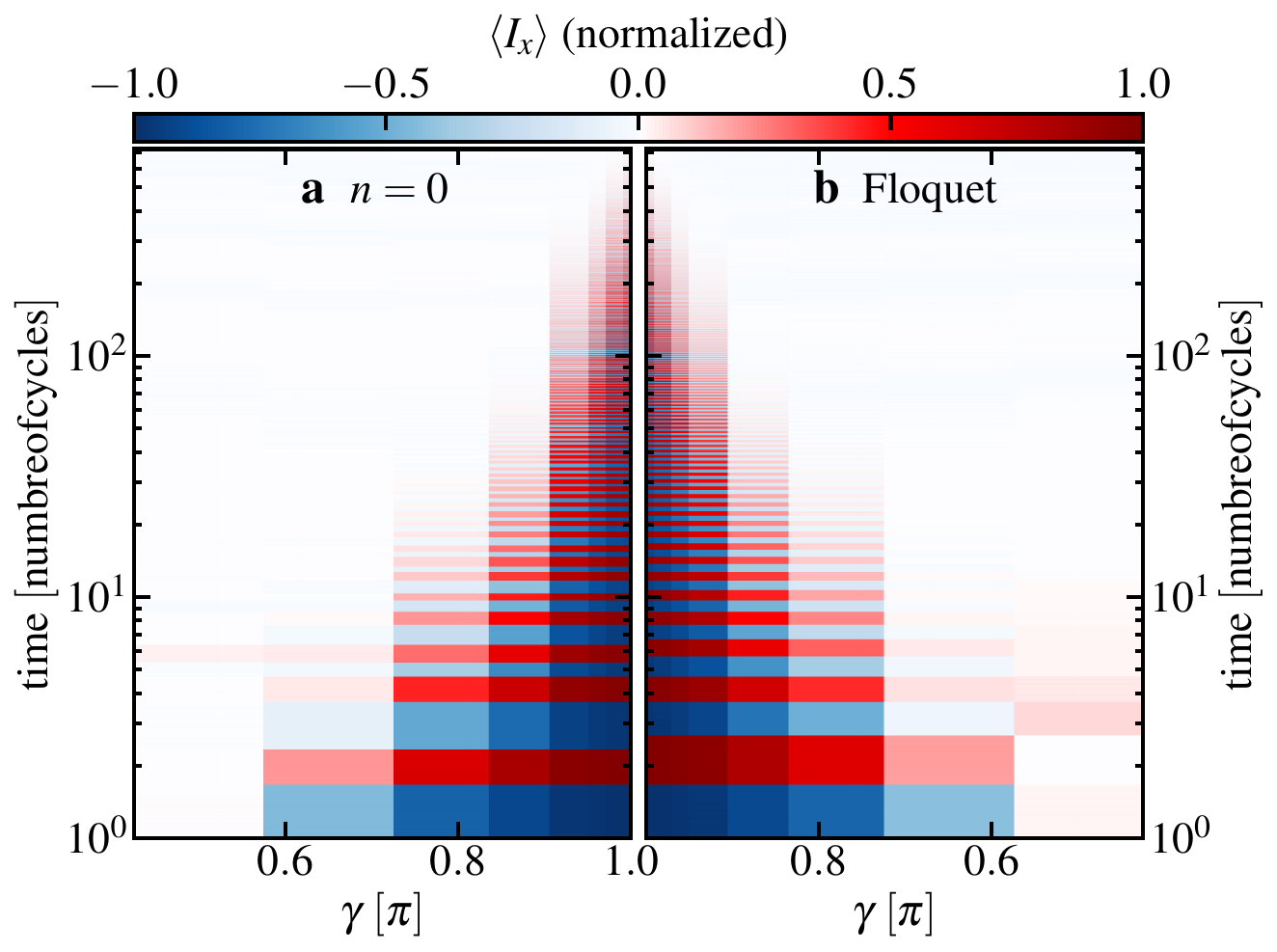}
		\caption{
			\textbf{Comparison Floquet and RMD-sequence:}
			Time evolution of polarization for different kick angles $\gamma$, for \textbf{a}, monopole~($n=0$) and \textbf{b}, Floquet sequence.
			There is little difference in the stability regime of the two orders. However, the RMD signal is non-periodic in contrast to the periodic Floquet DTC response.
			Parameters are as in Fig.~\ref{fig:phase_diagram}.
			%\mb{I'm wondering whether the mismatch becomes more pronounced on a normal (instead of a log) y-xis. Maybe we can add a zoom?}
		}
		\label{SI:fig:phase_diagram_realtime}
	\end{figure}

	\begin{figure}[t]
		\centering
		\includegraphics[width=0.5\textwidth]{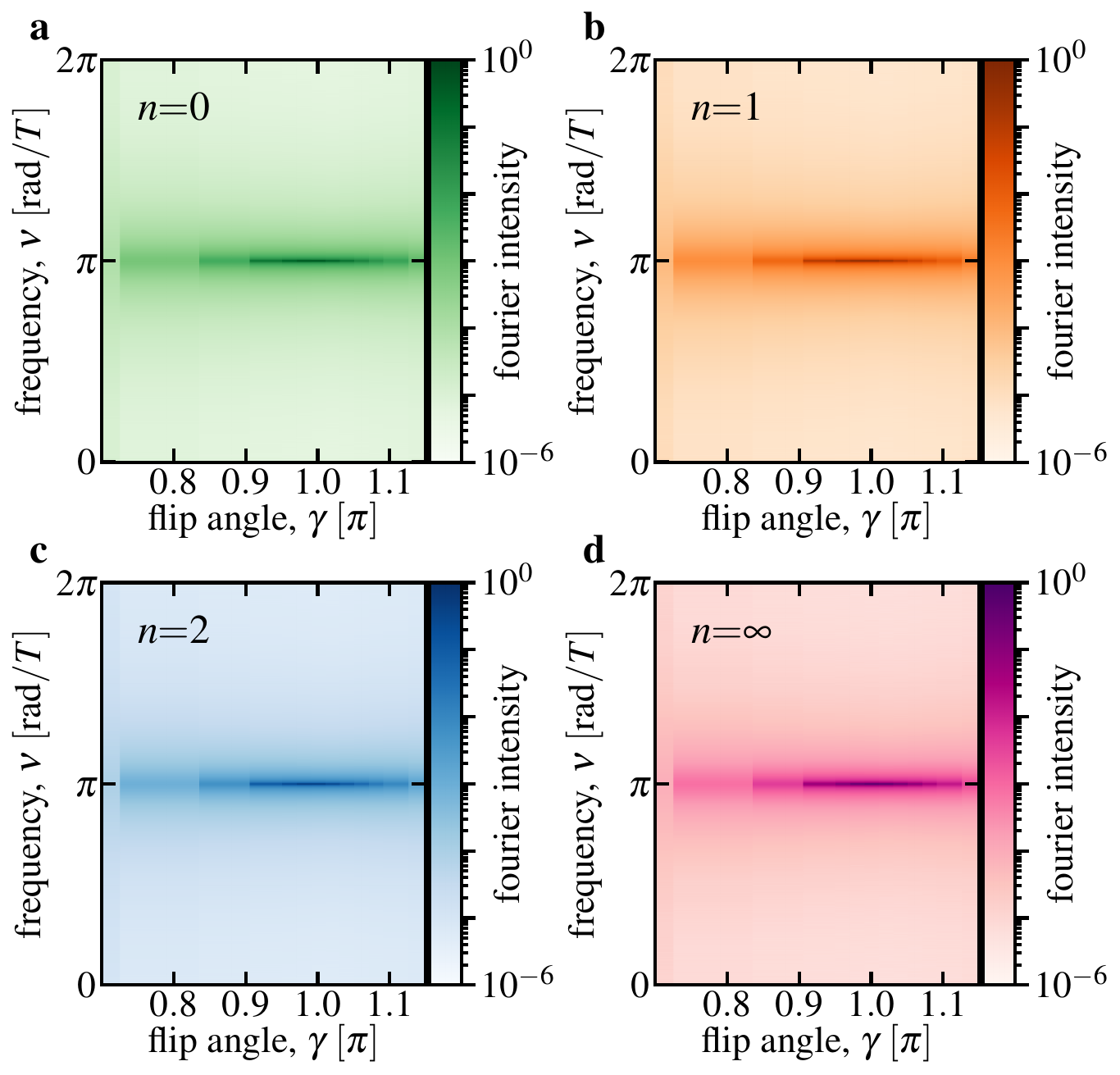}
		\caption{
			\textbf{Phase diagram for different multipole orders.}
			Intensity of Fourier transform---defined as the absolute value squared of the discrete Fourier transform~(DFT), $|\text{DFT}|^2$---of stroboscopic dynamics for monopole~($n=0$, \textbf{a}), dipole~($n=1$, \textbf{b}), quadrupole~($n=2$, \textbf{c}), and Thue-Morse sequence~($n=\infty$, \textbf{d}).
			No difference is observed between the different monopole orders, in agreement with the expectation that $n>0$ corresponds to an interpolation between $n=0$ and $n=\infty$.
			Parameters are as in main text Fig.~\ref{fig:phase_diagram}.
		}
		\label{SI:fig:phase_diagram_all}
	\end{figure}

	\section{Properties of prethermal temporal order}

	For the sake of completeness, let us mention that, in the regime $\gamma_y\approx\pi$, the observed dynamics for all RMD-sequences satisfy the required properties of prethermal temporal order:
	\begin{enumerate}
		\item \textit{Stable spatiotemporal symmetry breaking} as indicated by the long-lived period doubling dynamics in Fig.~\ref{fig:intro}d.
		\item \textit{Robustness to perturbations}, specifically, the persistence of temporal order for finite detuning~($\varepsilon=\gamma-\pi$), as demonstrated in Fig.~\ref{fig:phase_diagram}.
		\item \textit{Parametrically controlled long-lived prethermal lifetime} as demonstrated in Fig.~\ref{fig:heating_rate}b.
		\item \textit{Independence of initial state}: In Fig.~\ref{sup_fig:init_state}, we demonstrate that due to the quasi-conserved $I_x$-polarization, the temporal order is independent of the initial state, as long as the initial state has a finite $I_x$-polarization. 
		Specifically, we tune the effective temperature of the initial state by exposing the spins to evolution under dipole-dipole interactions only, for a finite waiting time~($t_d$). This leads to an exponential decay in the waiting time of the initial polarization. After this initial decay period we apply the RMD-sequence~(see Fig.~\ref{fig:intro}). We observe that the resulting $I_x$-polarization dynamics are independent of the initial decay time, up to an overall multiplicative factor stemming from the initial decay, see Fig.~\ref{sup_fig:init_state}.
	\end{enumerate}

	\begin{figure}[t]
		\centering
		\includegraphics[width=0.5\textwidth]{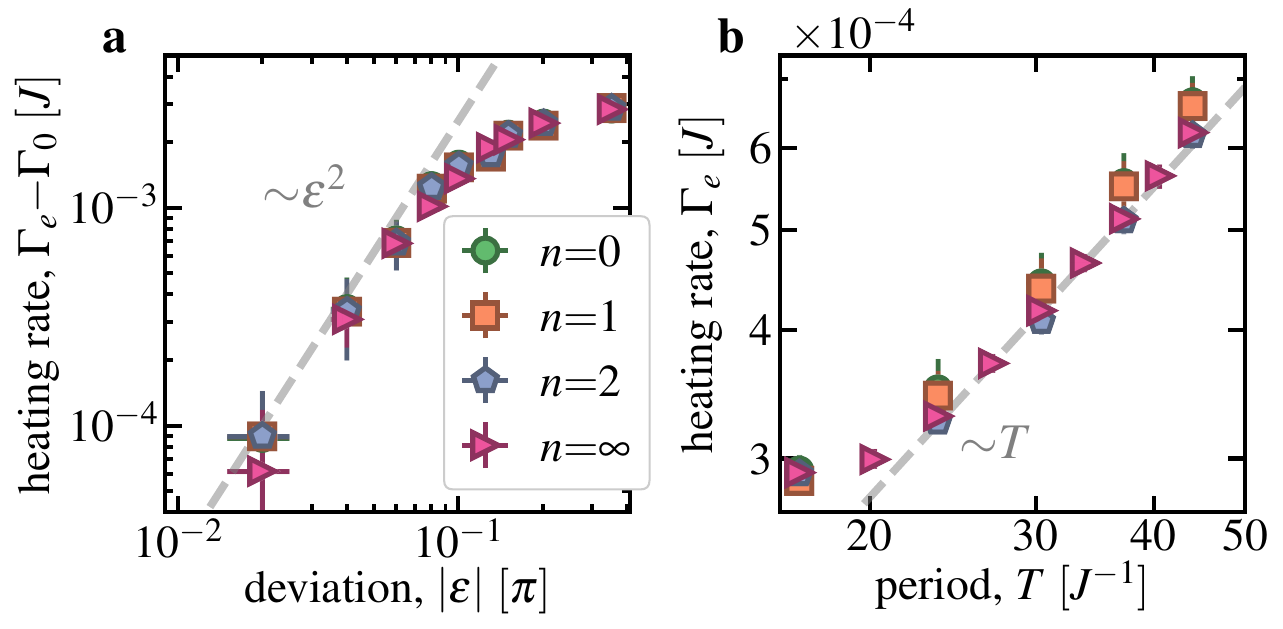}
		\caption{
			\textbf{Tunable lifetime for different multipole orders.}
			Same as main text Fig.~\ref{fig:heating_rate} for monopole~($n=0$), dipole~($n=1$), quadrupole~($n=2$) and Thue-Morse~($n=\infty$) RMD sequence. 
			\textbf{a}, Heating rate as a function of $y$-pulse deviation, $\varepsilon = \gamma_y - \pi$.
			\textbf{b}, Heating rate as a function of the period $T$, while simultaneously changing the detuning $\varepsilon=BT$.
			The heating rate in the observed parameter regime is independent of the multipole order $n$. Parameters are as in main text Fig.~\ref{fig:heating_rate}.
		}
		\label{SI:fig:lifetime_all}
	\end{figure}
	
	\section{Additional Data sets}
	
	\subsection{Direct Comparison with Floquet DTCs}
	
	In Fig.~\ref{SI:fig:phase_diagram_realtime}, we contrast the real-time evolution of a single-shot $0$-RMD sequence with a Floquet sequence at different kick-angles $\gamma$.
	While there is no difference in the stability of the observable polarization lifetime, the non-stroboscopic micromotion dynamics of the two sequences are clearly distinct: The periodicity of the Floquet drive is mirrored in the extracted signal; by contrast, the $0$-RMD sequence displays disordered non-stroboscopic dynamics.

	\subsection{Stability of Rondeau Order for Different Multipolar Drives}
	
	We briefly report on the stability of different multipole orders $n=0,1,2$ and Thue-Morse. 
	As indicated in the main text, we observe no difference in terms of stability or heating rate for the different multipole orders, see Fig.~\ref{SI:fig:phase_diagram_all} and Fig.~\ref{SI:fig:lifetime_all}, respectively. These results are consistent with the dephasing limit arguments above, which is independent of the multipole order $n$.

\end{document}